\documentclass[twocolumn,amsmath,amssymb,superscriptaddress,a4paper,accepted=2024-02-22]{quantumarticle}

\pdfoutput=1
\usepackage[numbers]{natbib}

\usepackage[skins,theorems]{tcolorbox}
\usepackage{tensor}
\usepackage{cancel}
\usepackage{color}
\usepackage{centernot}
\usepackage{graphicx}
\graphicspath{{./Figures/}} 
\usepackage{dcolumn}
\usepackage{bm}
\usepackage{bbold}
\usepackage{braket}
\usepackage{esdiff}
\usepackage{xr-hyper}
\usepackage{hyperref}
\usepackage[switch, mathlines,displaymath]{lineno}
\usepackage{subfigure}

\makeatletter
\newcommand*{\addFileDependency}[1]{
  \typeout{(#1)}
  \@addtofilelist{#1}
  \IfFileExists{#1}{}{\typeout{No file #1.}}
}
\makeatother

\begin{document}

\title{Entanglement dynamics of photon pairs and quantum memories in the gravitational field of the earth}

\author{Roy Barzel}
\email{roy.barzel@zarm.uni-bremen.de}
\affiliation{ZARM, University of Bremen, Am Fallturm 2, 28359 Bremen, Germany}
\author{Mustafa Gündoğan}
\affiliation{Institut f\"{u}r Physik, Humboldt-Universit\"{a}t zu Berlin, Newtonstraße 15, 12489 Berlin, Germany}
\affiliation{IRIS Adlershof, Humboldt-Universit\"{a}t zu Berlin, Zum Großen Windkanal 2, 12489 Berlin, Germany}
\author{Markus Krutzik}
\affiliation{Institut f\"{u}r Physik, Humboldt-Universit\"{a}t zu Berlin, Newtonstraße 15, 12489 Berlin, Germany}
\affiliation{IRIS Adlershof, Humboldt-Universit\"{a}t zu Berlin, Zum Großen Windkanal 2, 12489 Berlin, Germany}
\affiliation{Ferdinand-Braun-Institut (FBH), Gustav-Kirchoff-Str.4, 12489 Berlin, Germany}
\author{Dennis R\"atzel}
\email{dennis.raetzel@zarm.uni-bremen.de}
\affiliation{ZARM, University of Bremen, Am Fallturm 2, 28359 Bremen, Germany}
\affiliation{Institut f\"{u}r Physik, Humboldt-Universit\"{a}t zu Berlin, Newtonstraße 15, 12489 Berlin, Germany}
\author{Claus L\"ammerzahl}
\affiliation{ZARM, University of Bremen, Am Fallturm 2, 28359 Bremen, Germany}
\affiliation{Institute of Physics, Carl von Ossietzky University Oldenburg, Ammerländer Heerstr. 114-118, 26129 Oldenburg, Germany}

\begin{abstract}
We investigate the effect of entanglement dynamics due to gravity -- the basis of a mechanism of universal decoherence -- for photonic states and quantum memories in Mach-Zehnder and Hong-Ou-Mandel interferometry setups in the gravitational field of the earth. We show that chances are good to witness the effect with near-future technology in Hong-Ou-Mandel interferometry. This would represent an experimental test of theoretical modeling combining a multi-particle effect predicted by the quantum theory of light and an effect predicted by general relativity. Our article represents the first analysis of relativistic gravitational effects on space-based quantum memories which are expected to be an important ingredient for global quantum communication networks. 
\end{abstract}

\maketitle

\section{Introduction}
It has become one of the major problems of theoretical physics to understand the interplay between our most successful theories, quantum mechanics (QM) and general relativity (GR) \citep{feynman2018feynman}. A resolution of this problem can only be driven by experiments or observations at the interface of the two theories. In addition, the race in the development of space-based quantum technologies, where quantum resources are generated and probed locally \citep{Aveline2020,Lachmann2021} or are exchanged over thousands of kilometers through the inhomogeneous gravitational field of Earth \citep{1200, Xu2019, Gundogan2021, Sidhu2021, Lu2022}, fuels the need to understand the influence of general relativistic effects on quantum resources also from a practical point of view.  
 
A particular example of an interesting fundamental effect at the interface of quantum mechanics and general relativity is the generation of entanglement between the internal energy structure of a quantum system and its external (motional) degrees of freedom (DOFs) due to gravitational time dilation or redshift. These entanglement dynamics (EDs) due gravity have been proposed to be witnessed in atom interferometry \citep{Zych2011}, with single photons in Mach-Zehnder (MZ) interference \citep{Zych2012}, photon pairs in Hong-Ou-Mandel (HOM) interference \cite{Mohageg2021:deep} and phonons in Bose-Einstein condensates \cite{Edward_Bruschi_2014}. For the case of massive quantum systems that are in superposition states of their center of mass degree of freedom, EDs due to gravity were found to induce decoherence \citep{Pikovski2015,Pikovski_2017}, underlining their fundamental significance. 

\begin{figure}[t]\centering
  \includegraphics[width=0.90\linewidth]{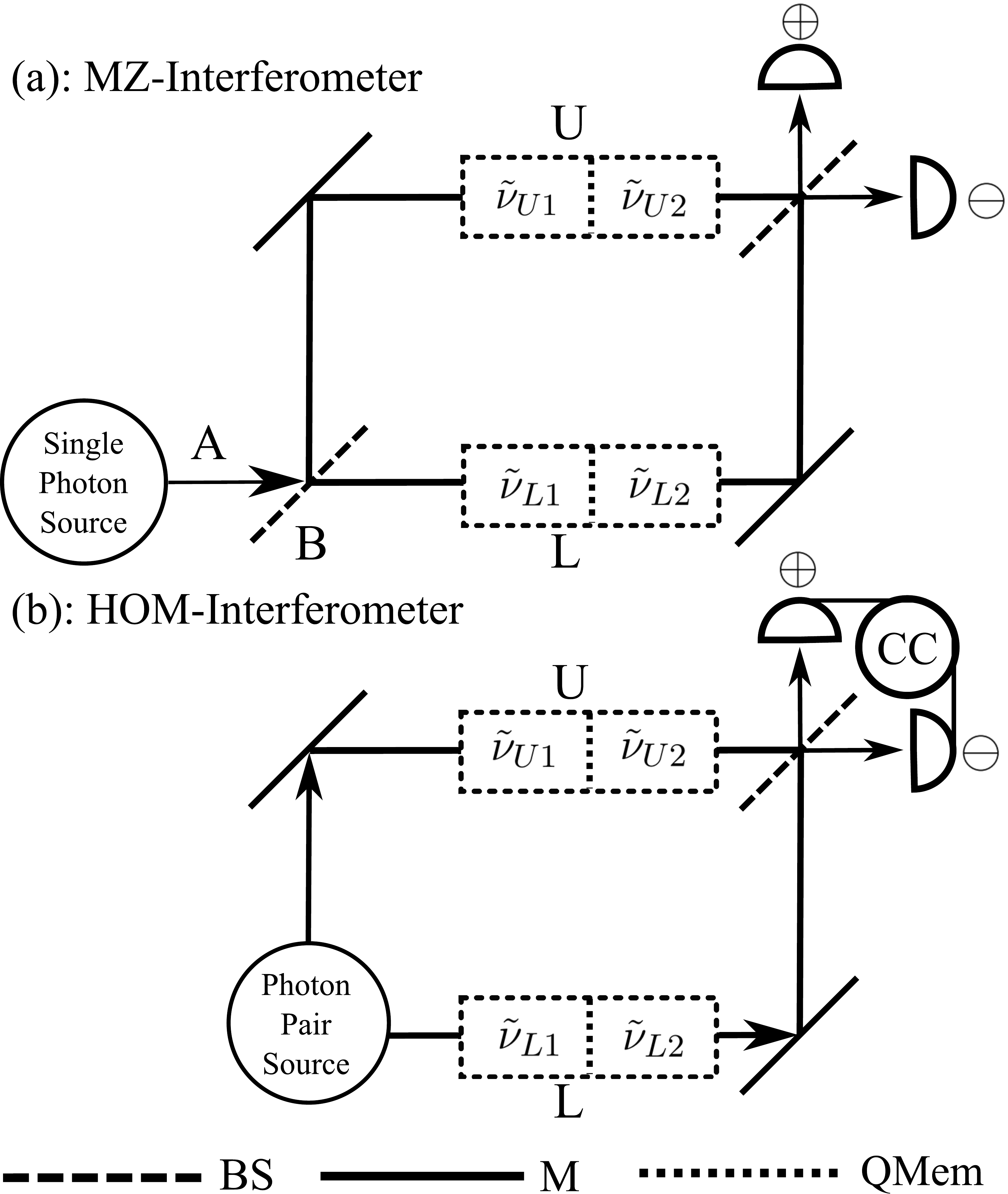}
  \caption{ Interference in gravitational field. (a): MZ-interferometer. (b): HOM-interferometer. BS: Beam splitter. M: Mirror. CC: Coincident count logic. QMem: Quantum memory. Each transition path is respectively equipped with two QMems with internal states, which \textit{locally} oscillate at discrete angular frequencies $\tilde\nu_{L i}$ and $\tilde\nu_{U i}$, i.e. the frequencies are measured locally in the rest frame of the respective QMem. They are tuned, together with the frequencies of corresponding control lasers, such that storage and readout process are optimized.
  }
\label{fig:HOM}
\end{figure}

In this article, we investigate the case of EDs of photons and Quantum Memories (QMems)~\citep{Afzelius2015, Heshami2015} due to gravity in MZ and HOM interferometry setups as shown in Fig. \ref{fig:HOM}. Furthermore, we provide an experimental proposal and a feasibility study to witness the effect in HOM experiments whose necessary spatial extensions are dramatically smaller than those of proposed experiments that only employ photons
\citep{Zych2012,pallister2017blueprint,Mohageg2021:deep}. Furthermore, we want to emphasize that 
the associated non-locality in the presence of entanglement between photons in different arms of a HOM interferometer extends the quality of the conceptional discussion from previous works considering MZ-interference of photons (see e.g. \cite{barzelPRD,bruschi2021gravitational,Mieling2022,Mohageg2021:deep} for other investigations of gravitational effects in interference of several particles).
Due to the close relation of EDs due to gravity to decoherence, our results are also of practical relevance as they inform us about potential new challenges to be faced in the development of novel space-based quantum technological applications.  
Before we derive the appearance of EDs of photons and QMems due to gravity in MZ- and HOM-interferometry setups, we first consider a simple example. In the following, we will set $\hbar=1$.

\section{Qubit in spatial superposition}
\label{sec:single_qubit}

The simplest system in which EDs due to gravity take place is a single two-level system, with internal energy states $\ket{\mu_{1,2}}$ of corresponding energies $\mu_{1,2}$ which is allowed to be situated at two different gravitational potential levels $U$ and $L$. In the following, we assume a stationary situation, where $U$ and $L$ correspond to two local laboratories that are at rest in Earth's gravitational field, where “at rest” means that they are located at a fixed position in the  co-rotating reference system of Earth, that is, geosynchronous and a fixed radial distance from Earth's center, i.e. they are assumed to move on timelike Killing trajectories through spacetime. For simplicity, we describe the dynamics of the considered system from the perspective of a third resting but not further specified reference observer, $o$, evolving w.r.t. to proper time $\tau_o$. Then, the considered two-level system evolves w.r.t. the proper time $\tau_{\sigma}=\Theta_\sigma \tau_{o}$ depending on the site $\sigma=U,L$, where $\Theta_\sigma=1+z_{{o}\sigma}$ is related to the redshift $z_{{o}\sigma}$ between the resting reference observer ${o}$ and the observers at $\sigma=U,L$, as defined in \citep{PhysRevD.95.104037}.  To leading order, we can describe the redshift with the  approximated formula \cite{philipp2020relativistic}
\begin{eqnarray}
    z_{{o}\sigma}&=&(W_\sigma-\nonumber W_{o})/c^2 \,,\quad \text{where}\\
    W_{\alpha}&=& -GM/R_{\alpha} - v_{\mathrm{orb},\alpha}^2/2\,,
\end{eqnarray}  
for $\alpha=\sigma,{o}$, and $R_{\alpha}$ and $v_{\mathrm{orb},\alpha}$ are the corresponding radial distance from the center of Earth and the orbital velocity, respectively, and $G$ is Newton's gravitational constant and $M$ is Earth's mass.

The single particle of our example can effectively be regarded as a two-qubit system, where one qubit is encoded in the  particle's internal DOF $\mu_{1,2}$ and the other qubit is encoded in its external DOF $\sigma$. The associated Hilbert space is of dimension four and is spanned by the basis  $\{\ket{\mu_1}\otimes\ket{U},\ket{\mu_2}\otimes\ket{U},\ket{\mu_1}\otimes\ket{L},\ket{\mu_2}\otimes\ket{L}\}$. 
To be subject to EDs due to gravity, the internal sub-state of the particle needs to describe a coherent superposition of energy eigenstates \begin{equation}
    \ket{\psi_\mathrm{int}(\tau_o)}=(\ket{\mu_1}+e^{i\varphi(\tau_{o})}\ket{\mu_2})/\sqrt{2}\,,
\end{equation}
usually denoted as clock state because during its time evolution it precesses like a clock around the equator of its associated Bloch-sphere. The relative phase $\varphi(\tau_{o})$ can then be identified as the time which the clock "shows".
 The single-particle Hamiltonian $\hat{H}_1$ which generates the time evolution of the particle is diagonal in the previously mentioned basis and reads 
 \begin{align}
     \hat{H}_1=\mathrm{diag}(\Theta_U\mu_1,\Theta_U\mu_2,\Theta_L\mu_1,\Theta_L\mu_2).\label{EQ:Hamilton1}
 \end{align}
 It can be deduced from a phenomenologically motivated ansatz, that is, effectively the coordinate time in the Schr\"odinger equation is replaced by the proper time \citep{Pikovski2015}.  
However, it reproduces the experimental results of the Hafele Keating experiment \citep{Hafele1972E} correct, in other words, one recovers the well known redshift relation of clock frequencies at different gravitational potential levels. EDs due to gravity take place for a clock-state in a coherent superposition of both sites 
\begin{equation}
    \ket{\phi(0)}=(\ket{\mu_1}+\ket{\mu_2})/\sqrt{2}\otimes(\ket{U}+\ket{L})/\sqrt{2}\,.
\end{equation}
This can be easily shown by first evolving the state with the single-particle Hamiltonian to $\ket{\phi(\tau_{o})}=e^{-i\hat{H}_1 \tau_{o} }\ket{\phi(0)}$, computing the corresponding density matrix $\hat{\rho}(\tau_{o})=\ket{\phi(\tau_{o})}\bra{\phi(\tau_{o})}$ and subsequently tracing out the internal (energy) DOFs of the particle, which results in the reduced $2\times2$ density matrix $\hat{\rho}_\mathrm{red}(\tau_{o})=\sum_{i=1,2}\braket{\mu_i|\hat{\rho}(\tau_{o})|\mu_i}$. 
The \textit{purity} of the reduced density matrix
\begin{equation}
    \mathcal{P}(\tau_{o})=\sum_{\sigma=U,L}\braket{\sigma|\hat{\rho}^2_\mathrm{red}(\tau_o)|\sigma}=1-\frac{\sin^2(\Delta_\Theta\mu_-\tau_{o}/2)}{2}\,,\label{EQ:1Q}
\end{equation}
with $\mu_-=\mu_1-\mu_2$ and $\Delta_\Theta:=\Theta_L-\Theta_U=z_{{o}L}-z_{{o}U}$ indicates that the initially pure spatial sub-state of the particle decohered into the maximally mixed state

\begin{equation}
    \hat{\rho}_\mathrm{red}(\tau_{{o},\mathrm{ent}}^{(1)})=(\ket{U}\bra{U}+\ket{L}\bra{L})/2
\end{equation} 
after the time
\begin{align}
\tau_{{o},\mathrm{ent}}^{(1)}=\frac{\pi}{\Delta_\Theta\mu_-}\label{EQ:tent}
\end{align} 
since $\mathcal{P}(\tau_{{o},\mathrm{ent}}^{(1)})=1/2$.
Because the entire state $\hat{\rho}(\tau_{o})$ is pure for all times, one can infer that the time evolution of $\hat{H}_1$ generated entanglement between the internal and positional DOFs of the considered particle. 

Equivalently, one can consider the \textit{linear entropy} of the reduced spatial sub-state
\begin{align}
    \mathcal{S}(\tau_{o})=1-\mathcal{P}(\tau_{o})=\frac{\sin^2(\Delta_\Theta\mu_-\tau_{o}/2)}{2},
\end{align}
as a measure of entanglement between the internal and external DOF of the considered particle. The maximal entanglement between external and internal DOFs at $\tau_{{o},\mathrm{ent}}^{(1)}$ does then correspond to a maximum of $\mathcal{S}$.   

\begin{table*}[t!]
    		\centering
            \begin{tabular}{lc} 
				\\ \hline
				  \textbf{Quantity}  &   \textbf{Symbol} \\ 
				\hline \hline 
                relative redshift & $z_{o\sigma}$   \\
                local redshift factor  &  $\Theta_\sigma$ \\
                difference frequencies &  $\mu_{-},\,\Omega_{-},\,\nu_{-}$   \\
                 sum frequencies  &   $\mu_{+},\,\Omega_{+},\,\nu_{+}$ \\
                differential redshift between $U$ and $L$ \quad & $\Delta_{\Theta}$ \\
                differential inverse redshift between $U$ and $L$ \quad & $\Delta_{\Theta^{-1}}$ \\
                purity of spatial substate &  $\mathcal{P}$    \\
                linear entropy of spatial substate &  $\mathcal{S}$   \\
                wavefunction normalization constants \quad & $N_\mathrm{MZ},\, N_\mathrm{HOM}$ \\
                local QMem storage times  & $\tau_{\sigma,s}$ \\ 
                interference patterns & $P_c^\mathrm{MZ},\, P_c^\mathrm{HOM}$ \\
                bunching anti-bunching probability difference & $P_c^\mathrm{HOM}$ 
            \end{tabular} 
		\caption{  Table of the most important recurring quantities and the corresponding symbols.
		\label{tab:symbols} }
\end{table*}

After having seen that gravitational effects can influence the entanglement between different DOFs of a single particle, it is natural to ask whether this effect might influence the mutual entanglement between different distant particles. The answer to this question is yes. In App. \ref{sec:A1}, we consider  the example of two particles subject to EDs due to gravity and show that some entanglement correlations between different particles feature non-trivial dynamics, where other entanglement correlations between different particles are robust against the gravitational effects. 

However, one should note that entanglement dynamics is a general result of the time evolution generated by a diagonal Hamiltonian of the form $\hat{H}_1$ of which the present example of EDs due to gravity is only a specific example.
The only relevant quantity is $\Delta_\Theta\mu_-=(d_1-d_2)-(d_3-d_4)$, where $d_i$ are the diagonal entries of $\hat{H}_1$. This quantity can be interpreted as the coupling between two qubits of a two-qubit system in general as it is a measure of how much the energetic distance between the two states of one qubit changes conditioned on the quantum state of the other qubit. Non-zero-values of $\Delta_\Theta\mu_-$ imply non-trivial entanglement dynamics in general. For instance, also a spin-$1/2$ particle which is exposed to magnetic fields of different magnitude at two different lattice sites $\sigma=U,L$ would experience the same effect, in complete absence of relativistic effects~\citep{Margalit2015}.

\section{EDs due to gravity in interference experiments}
 \label{sec:GIED}

Now, we draw our attention towards effects of EDs due to gravity in interference experiments with quantum states of photons involving quantum memories (QMems). We describe pure single- and two-photon states as
{\small
\begin{subequations}
\label{EQ:PS}
\begin{align}
\ket{\psi_{\mathrm{1}}}=&\sum_\sigma\int d\omega\, \psi_{\sigma}(\omega)\hat{a}^\dagger_{\sigma \omega}\ket{0},    \label{EQ:SPS}
\\
\ket{\psi_{\mathrm{2}}}=& \frac{1}{\sqrt{2}} \sum_{\sigma_1,\sigma_2}\int d\omega_1 d\omega_2\, \psi_{\sigma_1\sigma_2}(\omega_1,\omega_2)\hat{a}^\dagger_{\sigma_1 \omega_1}\hat{a}^\dagger_{\sigma_2 \omega_2}\ket{0},    \label{EQ:TPS}
\end{align}
\end{subequations}
}
where $\omega$ is the internal frequency/energy DOF and $\sigma$ is the spatial DOF of the considered photon. The operators $\hat{a}_{\sigma\omega}^\dagger$ create a photon of frequency $\omega$ in spatial mode $\sigma$ without knowledge of polarization, $\ket{0}$ is the corresponding vacuum state of the electromagnetic field and $\psi_{\sigma}(\omega)$ and $\psi_{\sigma_1\sigma_2}(\omega_1,\omega_2)$ are the \textit{photonic wave functions}.  

Typically MZ-interferometers
are operated with initial photonic wave function 
\begin{equation}
    \psi^{\mathrm{}}_{\sigma}(\omega)=\psi_{}(\omega)\delta_{\sigma A},
\end{equation}
where $\delta_{}$ is the Kronecker delta and $A$ is one of the two input ports (A,B) of the first beam splitter of the MZI (see Fig.\ref{fig:HOM} a)). We consider the HOM-interferometer to be operated with initial two-photon wave function
\begin{align}
    \psi^{\mathrm{}}_{\sigma_1\sigma_2}(\omega_1,\omega_2)=& \, \Big(\psi_{}(\omega_1,\omega_2)\delta_{\sigma_1 U}\delta_{\sigma_2 L} \nonumber \\
    & \quad + \psi_{}(\omega_2,\omega_1)\delta_{\sigma_2 U}\delta_{\sigma_1 L}  \Big)/\sqrt{2}
\end{align}
as displayed in Fig. \ref{fig:HOM} b). Note the symmetry of the two-particle photonic wave function under simultaneous exchange of $\omega$ and $\sigma$, that is, $\psi_{\sigma_1\sigma_2}(\omega_1,\omega_2)=\psi_{\sigma_2\sigma_1}(\omega_2,\omega_1)$, which incorporates the indistinguishability of the two particles. We are interested in single-photon coherent superposition states of two angular frequencies $\Omega_1$ and $\Omega_2$ (the formal analog to the single-qubit clock-states of our example above)
\begin{align}
\psi(\omega)={N}_{\mathrm{MZ}}[g_{\Omega_1\xi}(\omega)+e^{i\varphi}g_{\Omega_2\xi}(\omega)]\label{EQ:SpecMZ}
\end{align}
and frequency entangled two-photon states (the formal analog to the entangled two-qubit clock-states from Appendix \ref{sec:A1})
\begin{align}
\psi(\omega_1,\omega_2)& ={N}_{\mathrm{HOM}}[g_{\Omega_1\xi}(\omega_1)g_{\Omega_2\xi}(\omega_2)\nonumber \\
& +e^{i\varphi}g_{\Omega_2\xi}(\omega_1)g_{\Omega_1\xi}(\omega_2)],\label{EQ:SpecHOM}
\end{align}
where $g_{\Omega\xi}(\omega)=(2\pi\xi^2)^{-1/4}e^{-(\omega-\Omega)^2/(4\xi^2)}$ is a Gaussian distribution, $\xi$ is the single-photon bandwidth and ${N}_{\mathrm{MZ}}$ and ${N}_{\mathrm{HOM}}$ are normalization constants. We assume the wave functions to be normalized as $\int d\omega |\psi(\omega)|^2=1$ and $\int d\omega_1d\omega_2 |\psi(\omega_1,\omega_2)|^2=1$. This determines the normalization constants to be
\begin{subequations}
\begin{align}
{N}_{\mathrm{MZ}}&= (2[1+\cos(\varphi)e^{\Omega_-^2/(8\xi^2)}])^{-1/2}
\\
{N}_{\mathrm{HOM}}&= (2[1+\cos(\varphi)e^{\Omega_-^2/(4\xi^2)}])^{-1/2}\,,
\end{align}
\label{EQ:A_Norms}
\end{subequations}
where $\Omega_-=\Omega_2-\Omega_1$. The spectra and all frequencies $\Omega_i$ are measured in the local reference frame $o$ which we associate with the photon (pair) source.

The initial photon states are injected into the interferometers as displayed in Fig. \ref{fig:HOM}. In case of the MZ-interferometer, a first beam splitter is applied which transforms the photonic creation operators as $\hat{a}_{A\omega}\rightarrow (\hat{a}_{U\omega}-\hat{a}_{L\omega})/\sqrt{2}$ and $\hat{a}_{B\omega}\rightarrow (\hat{a}_{U\omega}+\hat{a}_{L\omega})/\sqrt{2}$. In both, the MZ- and HOM-interferometer, frequency-dependent phases $\varphi_\sigma(\omega)$ are added due to storage process and propagation. Then, $50:50$ beam splitters are applied, which transform the photonic creation operators as $\hat{a}_{U\omega}\rightarrow (\hat{a}_{\oplus\omega}-\hat{a}_{\ominus\omega})/\sqrt{2}$ and $\hat{a}_{L\omega}\rightarrow (\hat{a}_{\oplus\omega}+\hat{a}_{\ominus\omega})/\sqrt{2}$.  These transformations lead to the final photonic wave functions $\psi^\mathrm{f}_{\sigma}(\omega)$ and $\psi^\mathrm{f}_{\sigma_1\sigma_2}(\omega_1,\omega_2)$.

In conventional MZ- and HOM-interferometry, the detectors are frequency/energy-insensitive. Thus we are only interested in the spatial sub state of the final state, i.e. in the reduced single- and two-photon density matrices 
\begin{subequations}
\label{EQ:rhomat}
\begin{align}
\rho_{\sigma\bar{\sigma}}=&\int d\omega\, \psi^{\mathrm{f}}_{\sigma}(\omega)\psi^{\mathrm{f}}_{\bar{\sigma}}(\omega)^*,\label{EQ:rhomat1}
\\
\rho_{\sigma_1\sigma_2\bar{\sigma}_1\bar{\sigma}_2}=& \,
 \int d\omega_1 d\omega_2\, \psi^{\mathrm{f}}_{ \sigma_1\sigma_2}(\omega_1,\omega_2) \psi^{\mathrm{f}}_{ \bar{\sigma}_1\bar{\sigma}_2}(\omega_1,\omega_2)^* \label{EQ:rhomat2},
\end{align}
\end{subequations}

The detection statistics of MZ- and HOM-experiments is encoded in the diagonal entries of the reduced density matrices (\ref{EQ:rhomat}) as we have 
\begin{subequations}
\label{eq:diagentriesredrho}
\begin{align}\label{eq:Psigma}
    P_{\sigma'}^{\mathrm{MZ}}&=\int d\omega\,\braket{\hat{a}^\dagger_{\sigma'\omega}\hat{a}_{\sigma'\omega}}=\rho_{\sigma'\sigma'}\quad\mathrm{and}\\
    \label{eq:Psigma1sigma2}
   \nonumber  P_{\sigma_1'\sigma_2'}^{\mathrm{HOM}}&=\int d\omega_1 d\omega_2\,  \braket{\hat{a}^\dagger_{\sigma_1'\omega_1}\hat{a}^\dagger_{\sigma_2'\omega_2}\hat{a}_{\sigma_2'\omega_2}\hat{a}_{\sigma_1'\omega_1}}{/2}\\
    & =\rho_{\sigma_1'\sigma_2'\sigma_1'\sigma_2'} \,,
\end{align}
\end{subequations}
where $\sigma',\sigma_1',\sigma_2'=\oplus,\ominus$.
In MZ-interference, coherence can be studied by analyzing the difference between the detector intensities
\begin{align}\label{eq:PcMZ}
 P_c^{\mathrm{MZ}}&=P_\ominus^{\mathrm{MZ}} - P_\oplus^{\mathrm{MZ}}\,.
\end{align}
In HOM-interference a measure of spatial 
correlation between the two involved photons is the difference between the photon bunching and photon anti-bunching probability
\begin{equation}\label{eq:PcHOM}
    P_c^{\mathrm{HOM}}=(P_{\oplus\oplus}^{\mathrm{HOM}}+P_{\ominus\ominus}^{\mathrm{HOM}})-(P_{\oplus\ominus}^{\mathrm{HOM}}+P_{\ominus\oplus}^{\mathrm{HOM}})\,.
\end{equation} 

As explained in the qubit model above, the size of the gravitational effect will depend on the time the photons spend on the different potential levels U and L. In this article, we investigate two options for the maximization of this time: QMems and optical delay lines. As our focus lies on the former, we discuss it first to compare the results to the latter case later.

\subsection{EDs due to gravity in quantum memories}

{ In this section, we discuss the application of QMems to store the photons in the interferometer arms as shown in Fig. \ref{fig:HOM}.} The storage process consists in a transfer of the photonic state to a matter system which we model as a mode swapping operation (details can be found in the experimental part of this paper and in App. \ref{sec:QMemModel}). This process generically alters the energies of the spectral components of the quantum state according to the given energy level structure of the QMem. 
We assume that each peak of the spectral wave function is mapped to a quantum memory according to the mapping $\tilde\Omega_{\sigma i}\rightarrow \tilde\nu_{ \sigma i}$, where $\tilde\nu_{\sigma i}$ is the oscillation frequency of the $i$-th QMem at the potential level $\sigma=U,L$ and $\tilde\Omega_{\sigma i}$ is the frequency of the $i$-th peak of the spectral wave function as measured from the perspective of an observer at the potential level $\sigma=U,L$. Here{,} as in the following, we will denote frequencies in the local frames with a tilde and frequencies defined from the perspective of the global reference fixed to the frame of the photon source without a tilde. We assume that the time evolution of the state inside each QMem is governed by the simple unitary  $\hat{U}_{\sigma,i}=e^{-i \tilde\nu_{\sigma i} \tau_{\sigma\mathrm{s}}}$, where $\tau_{\sigma\mathrm{s}}$ is the local proper storage time at the potential level $\sigma=U,L$. 

\begin{figure}[h]\centering
\includegraphics[width=0.7\linewidth]{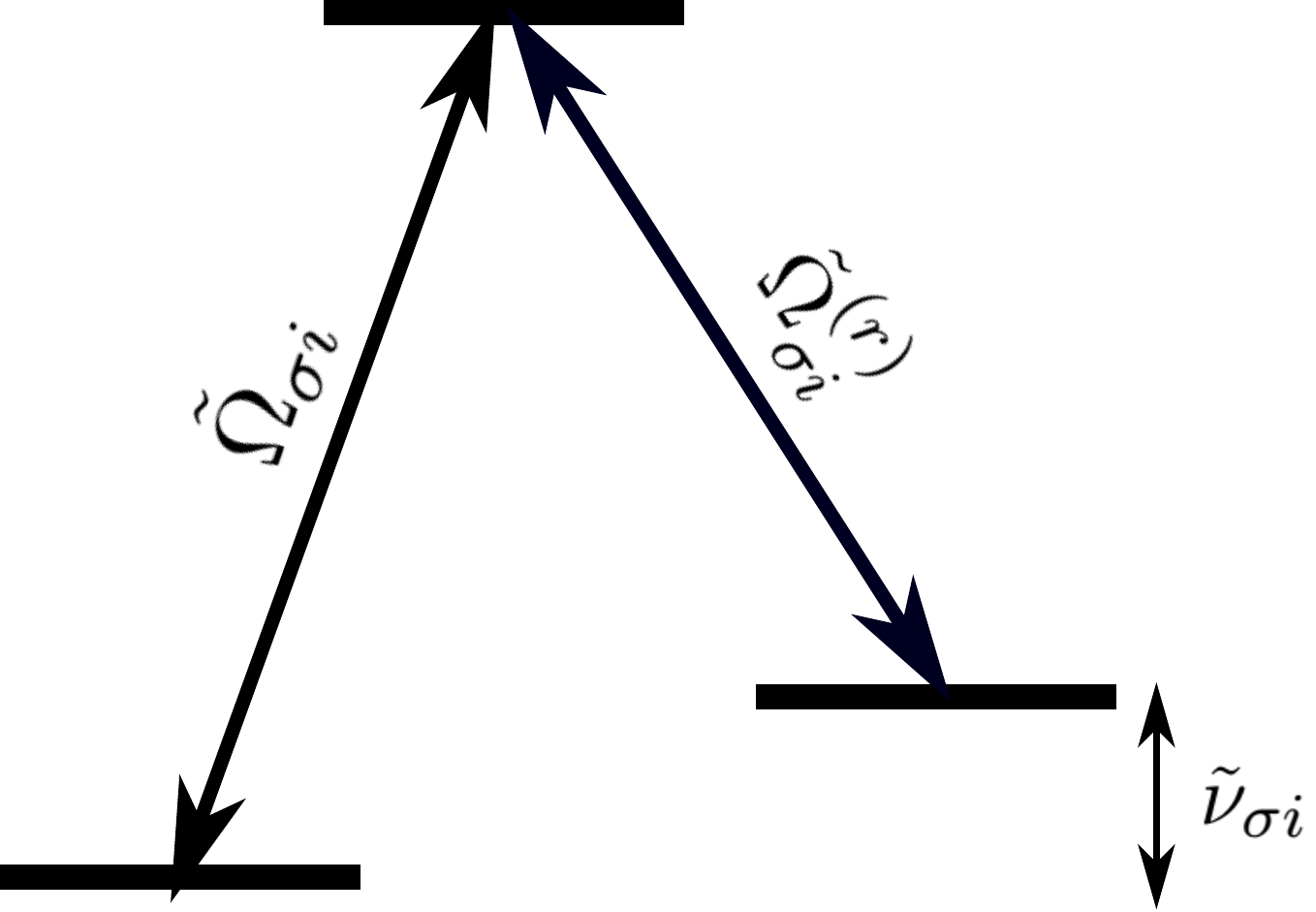}
  \caption{ Energy scheme for ground state QMems denoted as Lambda system. The two lower states are usually hyperfine sublevels of the same electronic state. The frequencies of the internal dynamics $\tilde{\nu}_{\sigma i}$ are then in the GHz regime. The acceptance-frequency of the $i$-th QMem is denoted as $\tilde{\Omega}_{\sigma i}$ and the frequency of the $i$-th control laser is $\tilde{\Omega}^{(r)}_{\sigma i}$. 
  }
  \label{fig:Lambda}
\end{figure}

After the storage time $\tau_{\sigma\mathrm{s}}$, the quantum state is read out from the memory and imprinted on a photonic state that is propagating to the beam splitter. The storage and read-out processes add additional phases due to the interaction with the external control lasers. These operate at the (typical optical) frequencies $\tilde\Omega_{\sigma i}^{(r)}$  in the local reference frames corresponding to $U$ and $L$. 
{ For the storage process to be effective, one generally has to assume a fixed relation between the local frequencies  $\tilde\nu_{\sigma i}$, $\tilde\Omega_{\sigma i}^{(r)}$ and $\tilde\Omega_{\sigma i}=\Omega_i/\Theta_\sigma$, where $\Theta_\sigma=1+z_{{o}\sigma}$ and $z_{{o}\sigma}$ is the redshift with respect to the photon source. In the following, we consider the case of a specific type of QMem called Lambda system (see Fig. \ref{fig:Lambda}, details can be found in Section \ref{sec:expimp}), where we set $\tilde\nu_{\sigma i} + \tilde\Omega_{\sigma i}^{(r)} = \tilde\Omega_{\sigma i}$.  }  
Performing a straightforward calculation (whose details we outline in App. \ref{sec:QMemModel}), the resulting state after the storage becomes
\begin{subequations}
\begin{align}
\ket{\psi_{\mathrm{MZ}}'}=& \sum_\sigma \int d\omega\, \psi_{\sigma}'(\omega)\hat{a}^\dagger_{\sigma\omega} \ket{0},
\\
\ket{\psi_{\mathrm{HOM}}'}=&\int d\omega_1 d\omega_2\, \psi'(\omega_1,\omega_2)a^\dagger_{U \omega_1}a^\dagger_{L \omega_2}\ket{0},\label{EQ:A_IN_PS2b}
\end{align}
\label{EQ:A_IN_PS2}
\end{subequations}
where the photonic wave functions altered by the QMems are
\begin{subequations}
\begin{align}
\nonumber \psi_{\sigma}'(\omega)=&\, (-1)^{\tilde{n}(\sigma)}N_{\mathrm{MZ}}[g_{\Omega_1\xi}(\omega) e^{i\tilde\Omega_{\sigma 1}\tau_{\sigma\mathrm{s}}} \\
& \quad + e^{i\varphi}g_{\Omega_2\xi}(\omega)e^{i\tilde\Omega_{\sigma 2}\tau_\mathrm{\sigma s}} ]/\sqrt{2}
\\
\nonumber \psi'(\omega_1,\omega_2)=&\, N_{\mathrm{HOM}}[g_{\Omega_1\xi}(\omega_1)g_{\Omega_2\xi}(\omega_2)\\
& \quad +e^{i\varphi'_{\mathrm{HOM}}}g_{\Omega_2\xi}(\omega_1)g_{\Omega_1\xi}(\omega_2)]\,,\label{EQ:APsiHOM}
\end{align}
\label{EQ:APsi}
\end{subequations}
$\tilde{n}(U)=0$, $\tilde{n}(L)=1$ and
\begin{align}
    \varphi'_{\mathrm{HOM}}=& \varphi + \tilde\Omega_{U-}\tau_{U\mathrm{s}} - \tilde\Omega_{L-}\tau_{L\mathrm{s}} \,
\end{align}
where $\tilde\Omega_{\sigma -} = \tilde\Omega_{\sigma 2}-\tilde\Omega_{\sigma 1} = (\Omega_{ 2}-\Omega_{1})/\Theta_\sigma$. Note that neither $\tilde\nu_{\sigma i}$ nor $\tilde\Omega_{\sigma i}^{(r)}$ appear in equations \eqref{EQ:APsi}. This is because the internal dynamics with $\tilde\nu_{\sigma i}$ and the temporal phase evolution of the control laser combine to the phase evolution with $\tilde\Omega_{\sigma i}$. The situation would be different, for example, if the internal evolution is not imprinted on the output state. The general results can be found in App. \ref{sec:QMemModel}. It should be noted, however, that for ground state QMems, the resulting phases due to the internal evolution are very small in comparison to the much larger phases imprinted by the control lasers (see also \ref{sec:expimp}).

Finally, to predict the outcome of the interference experiments shown in Fig. \ref{fig:HOM}, we have to specify the relation of $\tau_{U\mathrm{s}}$ and $\tau_{L\mathrm{s}}$. There are two basic options for the synchronization of the storage processes: equivalent local storage times and synchronization in a global reference frame. In the latter case, $\tau_{Us}/\Theta_U=\tau_{Ls}/\Theta_L$ and no gravitational effect on the interference patterns can be observed.

The first case comprises to set $\tau_{U\mathrm{s}}=\tau_{L\mathrm{s}}=\tau_\mathrm{s}$. For example, this may be achieved by placing a reference clock next to each of the QMems. 
We obtain for the coherence measures (\ref{eq:PcMZ}) and (\ref{eq:PcHOM}) in the large bandwidth limit (where expressions simplify significantly and analytical calculations can be performed with reasonable effort)
$\Omega_2-\Omega_1\gg \xi$ the MZ- and HOM-interference patterns
\begin{subequations}
\label{EQ:IP_}
\begin{align}
P_c^{\mathrm{MZ}}=&\cos(\Delta_{\Theta^{-1}} \Omega_+ \tau_\mathrm{s}/2) \cos(\Delta_{\Theta^{-1}} \Omega_- \tau_\mathrm{s}/2),\label{EQ:IPMZ}
\\
P_c^{\mathrm{HOM}}=&\cos(\Delta_{\Theta^{-1}} \Omega_- \tau_\mathrm{s} - \varphi),\label{EQ:IPHOM}
\end{align}
\end{subequations}
where $\Omega_\pm=\Omega_2\pm\Omega_1$ and $\Delta_{\Theta^{-1}}=1/\Theta_L - 1/\Theta_U$. We want to emphasize the appearance of $\Delta_{\Theta^{-1}}$ in the frequencies of EDs in contrast to the qubit case discussed above, where the frequency of EDs is proportional to $\Delta_{\Theta}$. This shows a subtle difference in the appearance of EDs due to gravity in these two cases: In the photon case, the effect is a result of gravitational redshift, and in the qubit case, it is gravitational time dilation. These effects are of course intimately related.  

From Eq. (\ref{EQ:IPMZ}), one can see that the  well known rapid oscillations of the MZ-interference signal are recovered in the first factor featuring the average frequency $\Omega_+/2$.
The rapid oscillations are modulated on a much longer time scale by the second factor featuring the difference frequencies. 
This modulation of the interferometric contrast is a result of EDs due to gravity. In other words, the internal frequency degree of freedom becomes entangled with the photon path which means that a measurement of the internal degree of freedom would provide which-path information, thereby erasing the interferometric contrast.
A full loss of contrast is obtained for $\Delta_{\Theta^{-1}} \Omega_- \tau_\mathrm{s}=\pi$, which corresponds to a storage time
\begin{equation}
    \tau^{{\mathrm{MZ}}}_\mathrm{s{,ent}}=\frac{\pi}{\Delta_{\Theta^{-1}}\Omega_-}\,.\label{EQ:tMZ}
\end{equation}

Indeed, with the help of Eq. (\ref{EQ:rhomat1}), one finds the reduced density matrix of the photon's spatial DOF for this setting in a maximally mixed state of both detector modes (c.f.  (\ref{EQ:AMZx})) 
\begin{align}
    \rho_{\sigma\bar{\sigma}}(\tau^{{\mathrm{MZ}}}_\mathrm{s{,ent}})=
    \frac{1}{2}\begin{pmatrix}
        1 & 0 \\
        0 & 1
    \end{pmatrix},
\end{align}
that attains the minimum value of one half for the purity (see Appendix \ref{sec:MixDecoh} for further details)
\begin{align}
    \label{EQ:LEMZ}\mathcal{P}_\mathrm{MZ}=\sum_{\sigma,\bar\sigma}|\rho_{\sigma\bar{\sigma}}|^2=\frac{1}{2}\left(1+\mathcal{V}^2\right),
\end{align}
where
\begin{align}
    \mathcal{V}=\cos(\Delta_{\Theta^{-1}} \Omega_- \tau_\mathrm{s}/2),
\end{align}
is the slowly oscillating envelope of the MZ-interferogramm (\ref{EQ:IPMZ}), known as the \textit{visibility}, the interferometric contrast. From Eq. (\ref{EQ:LEMZ}) it is apparent that lower values of the interferometric contrast $\mathcal{V}$ correspond to lower values of the purity of the reduced spatial sub-state. Therefore, the loss of the interferometric contrast, i.e. $\mathcal{V}=0$, maximizes the linear entropy
\begin{align}
    \label{EQ:SSMZ}\mathcal{S}_\mathrm{MZ}=1-\mathcal{P}_\mathrm{MZ}=\frac{1}{2}\left(1-\mathcal{V}^2\right),
\end{align}
which also in this case serves as a measure of entanglement between the internal and external DOFs of the considered photon.

The interference pattern of the HOM-experiment (\ref{EQ:IPHOM}) vanishes for $\Delta_{\Theta^{-1}}\Omega_- \tau_\mathrm{s} =\pi/2$, that is, $P_c^{\mathrm{HOM}}=0$, which means that photon bunching and photon anti-bunching are equiprobable at this setting, i.e. the photons are not correlated (entangled) in their spatial DOF. Also, here the physical explanation is that during the propagation through the HOM-interferometer, the frequency DOF of each of the two entangled photons respectively gets entangled with the path of the respective photon. Put differently, the theorem of entanglement monogamy implies that the frequency-path entanglement can only be generated at the cost of lowering the 
entanglement between the photons' spatial DOFs \citep{barzel2022role}, reflected by the equiprobability of photon bunching and photon anti-bunching. 
Indeed, with the help of Eq. (\ref{EQ:rhomat2}), one finds the spatial substate of the photons at the corresponding storage time  
\begin{align}
    \tau^\mathrm{HOM}_\mathrm{s,ent}=\frac{\pi}{2\Delta_{\Theta^{-1}}\Omega_-}\label{EQ:tHOM}
\end{align}

 to be (c.f. Eq. (\ref{EQ:Add}))
\begin{align}
\rho_{\sigma_1\sigma_2\bar{\sigma}_1\bar{\sigma}_2}\label{EQ:RDHOM}(\tau^\mathrm{HOM}_\mathrm{s,ent})
    =\frac{1}{4}\begin{pmatrix}
        1 & 0 & 0 & -1\\
        0 & 1 & -1 & 0\\
        0 & -1 & 1 & 0\\
        -1 & 0 & 0 & 1
    \end{pmatrix},
\end{align}
where we have chosen for the row and column numbering $\{\oplus\oplus,\oplus\ominus,\ominus\oplus,\ominus\ominus\}$.
The reduced spatial density matrix (\ref{EQ:RDHOM}) characterizes a mixed state as certified by the 
purity (see Appendix \ref{sec:MixDecoh} for further details)
\begin{align}
    \mathcal{P}_{\mathrm{HOM}}=\sum_{\substack{\sigma_1,\sigma_2, \\ \bar\sigma_1,\bar\sigma_2}}|\rho_{\sigma_1\sigma_2\bar{\sigma}_1\bar{\sigma}_2}|^2=\frac{1}{2}\left(1+\left(P_c^{\mathrm{HOM}}\right)^2\right)
\end{align}
as one in particular finds
\begin{align}
    \label{EQ:PHOM0}\mathcal{P}_\mathrm{HOM}(\tau^\mathrm{HOM}_\mathrm{s,ent})=\frac{1}{2}.
\end{align}
Also in the HOM-experiment, the linear entropy of the spatial sub-state
\begin{align}
    \label{EQ:LEHOM}\mathcal{S}_{\mathrm{HOM}}=1-\mathcal{P}_{\mathrm{HOM}}=\frac{1}{2}\left(1-\left(P_c^{\mathrm{HOM}}\right)^2\right)
\end{align}
is a measure of entanglement between the internal and external DOFs of the photons, which is maximized at the equiprobability of photon bunching and photon anti-bunching, i.e. $P_c^{\mathrm{HOM}}=0$. 

Not only entanglement between the internal and external DOFs of the photons is built up in HOM-interference. At the same time, the mutual entanglement shared between the spatial degrees of freedom of the photons and the spectral degrees of freedom of the photons, respectively, is depleted. We discuss the mutual entanglement between the photons only in their spatial DOF in the following, but want to emphasize that completely analogous results can be derived, in principle, for their spectral DOF (although the entanglement analysis and the associated continuous eigenvalue problem is slightly more involved). As $P_c^{\mathrm{HOM}}$ quantifies the spatial correlation of the photons to exit the same or distinct output ports of the beam splitter in the HOM-experiment, this quantity is an intuitive candidate for a measure of spatial entanglement shared between the photons. This is indeed the case as the absolute value of $P_c^{\mathrm{HOM}}$ is proportional to the \textit{negativity}, i.e.
\begin{align}
    \mathcal{N}_\mathrm{HOM}=&-\sum_{\lambda_{\mathrm{PT}}<0} \lambda_{\mathrm{PT}}=\frac{|P_c^{\mathrm{HOM}}|}{2},
\end{align}
where $\lambda_{\mathrm{PT}}$ are the eigenvalues of the partial transpose of the reduced spatial density matrix \eqref{EQ:rhomat2}. 
Apart from being an entanglement monotone, the negativity unambiguously certifies by its positivity the presence of entanglement in the spatial sub-state of the two photons considered here (see Appendix \ref{sec:MixDecoh} for derivation and discussion). Thus, Eq. (\ref{EQ:LEHOM}) asserts that the creation of entanglement between the internal and external DOFs of the involved photons  (i.e. the increase of $\mathcal{S}_{\mathrm{HOM}}$) comes at the cost of depleting the mutual spatial entanglement shared between the photons (i.e. the decrease of ${P}_c^{\mathrm{HOM}}$), and therefore in turn emphasizes the role of entanglement monogamy in HOM-interference.

Note that the mutual spatial entanglement between the two photons in the HOM-experiment is depleted twice as fast as the entanglement between the internal and external DOFs of a single photon in MZ-interference is  
built up. Further note that, in contrast to the MZ-interference pattern, the HOM-interference pattern is not subject to high-frequency oscillations with periodicity $\Omega_+/2$ 
which makes the HOM-experiments more robust against finite detector resolution and noise effects. However, also in the HOM-case one has to impose 
temporal resolution limits on the employed optical delays and the storage times of the QMems as $\tau^\mathrm{reslim}_\mathrm{s}=\pi/4\Omega_-$ in order to resolve the oscillations of the interference patterns which would otherwise be washed out by phase fluctuations.

Note also that we want to observe at least a half cycle of the entanglement dynamics such that the spatial correlations are not just lost but a full revival is witnessed. This is because, the loss of spatial correlations could also be, for example, due to decoherence through interaction with the environment.

\subsection{EDs due to gravity in delay lines}

If one wants to witness EDs due to gravity just with free photons, they have to travel for distances of the order of $c\tau_{s,\mathrm{ent}}$. It has been proposed to employ optical delay lines for this purpose (see e.g. \cite{Mohageg2021:deep}).
Applying locally equivalent optical delay lines in the distinct interferometer arms with local delay time $\tau_d$ (given by the length divided by the speed of light in the fiber) leads to a time evolution of the photonic state, which is governed by the unitary evolution $\hat{a}_{\sigma\omega}\rightarrow \hat{a}_{\sigma\omega}e^{i\omega \tau_\mathrm{d}/\Theta_\sigma}$.
In this case, we obtain for the spectra (\ref{EQ:SpecMZ}) and (\ref{EQ:SpecHOM}) in the large bandwidth limit 
$\Omega_2-\Omega_1\gg \xi$ and 
$\Delta_{\Theta^{-1}} \tau_\mathrm{d} \xi\ll 1$ the MZ- and HOM-interference patterns (see App. \ref{sec:dervPat} for the derivation)
\begin{subequations}
\label{EQ:IP_d2}
\begin{align}
P_c^{\mathrm{MZ}}=&\cos(\Delta_{\Theta^{-1}} \Omega_+ \tau_\mathrm{d}/2) \cos(\Delta_{\Theta^{-1}} \Omega_- \tau_\mathrm{d}/2),\label{EQ:IPMZ2}
\\
P_c^{\mathrm{HOM}}=&\cos(\Delta_{\Theta^{-1}} \Omega_- \tau_\mathrm{d} - \varphi),\label{EQ:IPHOM2}
\end{align}
\end{subequations}
The equations \eqref{EQ:IP_d2} are equivalent to equations \eqref{EQ:IP_} up to the replacement $\tau_\mathrm{s}\rightarrow \tau_\mathrm{d}$. Accordingly, the discussion following equations \eqref{EQ:IP_} equivalently applies to the case of delay lines under the replacement of the storage time by the optical delay time.

As for the case of the QMems (with equivalent local storage times), the effect of the delay lines arises due to the gravitational redshift of the photons and the resulting shift of the frequency difference between the two components of the photons' spectral wave function.
Alternatively \footnote{{By performing the temporal Fourier transform of the spectral wave function after applying the frequency dependent phase shift operation corresponding to the optical delay line, one obtains photon wave packets that are temporally translated by $\Theta_\sigma \tau_d$.}}, the effect of the delay lines can be interpreted in terms of the Shapiro delay \cite{Zych2012}, that is the variation of the speed of light depending on the gravitational potential as perceived from the perspective of a distant observer \cite{Shapiro1964,Shapiro1971}.

\section{Experimental implementation}
\label{sec:expimp}

\begin{figure}[t!]\centering
\vspace{0.5cm}
  \includegraphics[width=0.95\linewidth]{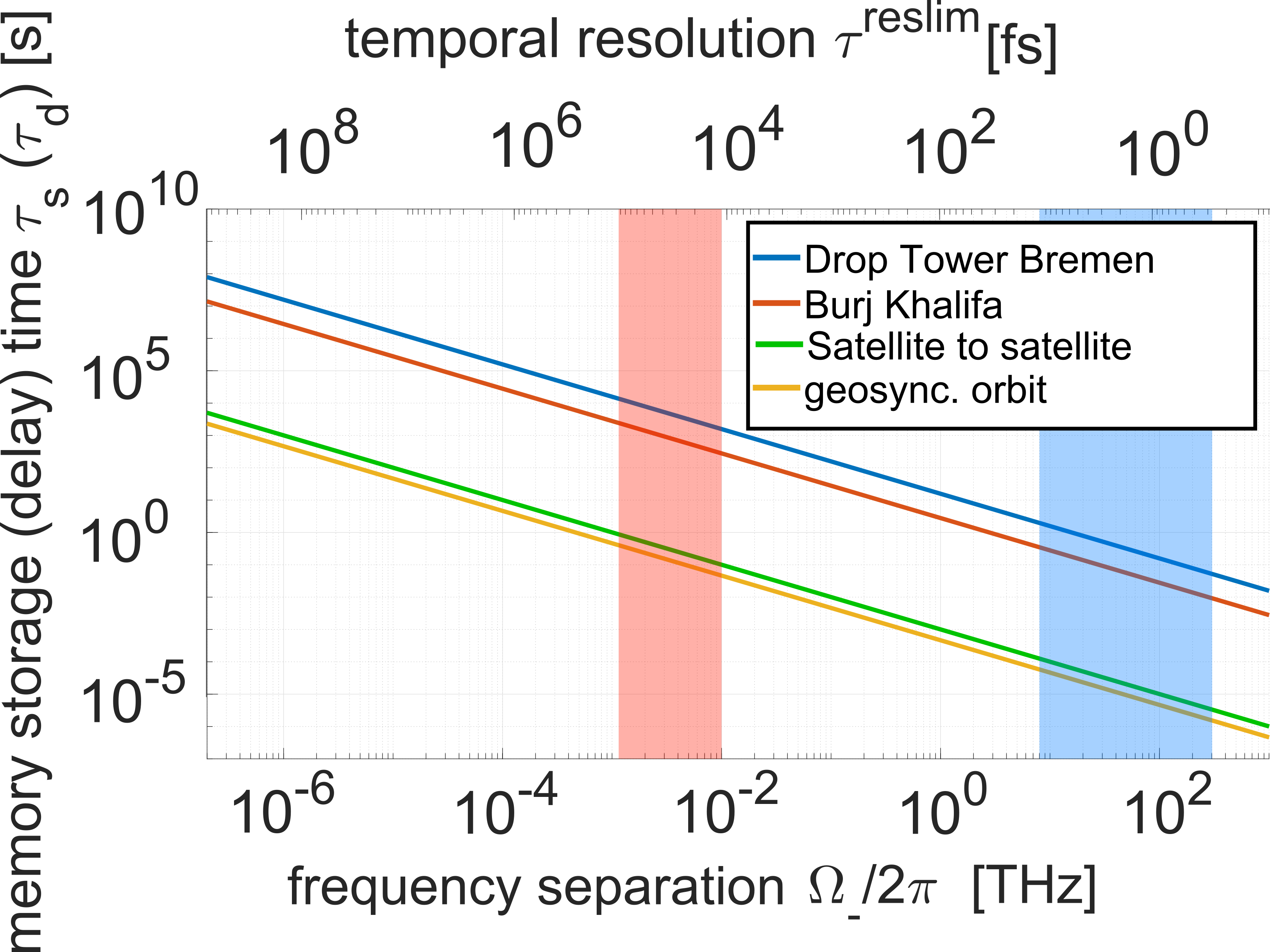}
  \caption{  
  Experimental requirements to witness EDs due to gravity in HOM-interference. The diagonal lines correspond to the required storage (delay) time $\tau_{\mathrm{s (d)}} = \pi/(2\Delta_{\Theta^{-1}} \Omega_-)$ in the HOM-experiment to observe EDs due to gravity for four different redshift factors: 
  Drop Tower Bremen ($\Delta_{\Theta^{-1}}\approx 1.6\cdot 10^{-14}$), Burj Khalifa ($\Delta_{\Theta^{-1}}\approx 9.0\cdot 10^{-14}$), satellite-to-satellite scenario ($\Delta_{\Theta^{-1}}\approx 2.5\cdot 10^{-10}$) and the redshift between a geostation and a geosynchronous orbit ($\Delta_{\Theta^{-1}}\approx 5.4\cdot 10^{-10}$). The red (blue) shaded area indicates the regime of frequency separation $\Omega_-=1-10\,\mathrm{GHz}$ ($\Omega_-=10-100\,\mathrm{THz}$) at which combinations of RIED QMems discussed in the main text operate. The upper horizontal axis indicates the required temporal resolution of the memory time (optical delay time) at a given frequency separation $\Omega_-$.  See  App. \ref{sec:estimation} for further details on the calculation of the required temporal resolution $\tau^\mathrm{reslim}$.}
\label{fig:times}
\end{figure}

Now we will discuss the experimental requirements to realize EDs due to gravity in our proposal. Firstly, single-photon frequency superposition states and frequency entangled two-photon states can be created via nonlinear optical methods such as spontaneous parametric down conversion~\citep{Rielander2017}, sum/difference frequency generation~\citep{Maring2017} and Bragg scattering four wave mixing~\citep{Clemmen2016}. The central element of our proposal are QMems with sufficiently long storage times that are capable of storing photonic states in the forms of Eqs. (\ref{EQ:SpecMZ}) and (\ref{EQ:SpecHOM}). Here one crucial requirement is that each QMem in Fig.~\ref{fig:HOM} should be able to store different frequency components, ($\tilde\Omega_{\sigma 1}$ and $\tilde\Omega_{\sigma 2}$) of the frequency superposition states. 
In Fig.~\ref{fig:times} we plot the required memory time, $\tau_{s}$ as a function of $\Omega_-$ for different arrangements of the levels $U$ and $L$.

Two different QMems operating at frequencies $\tilde\Omega_{\sigma 1}$ and $\tilde\Omega_{\sigma 2}$ can be combined together in each arm of the interferometer to ensure the efficient storage of both frequency components. We identified two different domains for such combinations: i) $\Omega_-\sim1-10$~GHz (red shaded area in Fig.\ref{fig:times}) and ii) $\Omega_-\sim10-200$~THz (blue shaded area). The first domain requires $\tau_{s}> 1\,$~h ($0.1-1$~s) for terrestrial (space-borne) implementations, whereas the second domain requires $\tau_{s} <1\,$~s ($10^{-5}$~s) for terrestrial (space-borne) experiments. Storage times of up to hours~\cite{Zhong2015, Ma2021} is within reach with rare-earth ion doped (REID) QMems~\citep{Gundogan2015, Jobez2015, Ortu2022}. REID systems owe such long storage times to their electronic structure: their optically active electronic orbitals lie within the filled outer electronic shells that create a shielding effect that results in long optical and spin coherence lifetimes~\cite{Goldner2015}. Furthermore, these system exhibit large inhomogeneous broadening which can be up to 20~GHz. This would allow storing different $\tilde\Omega_{\sigma 1}$ and $\tilde\Omega_{\sigma 2}$ components with equally long storage times within the same material through frequency multiplexing~\cite{Seri2019, Fossati2020}. For this scenario the required temporal resolution, $\tau^\mathrm{reslim}$, is around $10-100$~ps which can be achieved with today's electronics. \textcolor{black}{Furthermore, synchronization between remote clocks that are separated farther then our proposal requires for terrestrial experiments have been achieved with near-femtosecond precision~\citep{Deschenes2016,Bergeron2019}. Picosecond precision of synchronization has been achieved in quantum optics experiments~\cite{Quan2016, Valivarthi2022}.} Thus, the first domain is fully compatible with the state-of-the-art in REID QMems. 

{$\Omega_-\sim10-100$~THz of the second domain can be achieved by combining different types of QMems while satisfying the requirements on $\tau_s$. For instance, Pr ($\tilde\Omega_{\sigma 1}=606$~nm)~\cite{Gundogan2015, Seri2019} and Eu ($\tilde\Omega_{\sigma 2}=580$~nm)~\cite{Jobez2015, Ma2021, Ortu2022} based REID QMems would yield $\Omega_-=22$~THz. Cold and warm alkali gases can also be considered for our scheme: optical memories with lifetimes beyond 1~s have been demonstrated with Rb~\cite{Dudin2013, Yang2016} ($\tilde\Omega_{\sigma 1}=795$~nm) and Cs~\cite{Katz2018} ($\tilde\Omega_{\sigma 2}=894$~nm) gases. This combination would result in $\Omega_-=42$~THz. The different frequency separations, which can be attained, are summarized in Tab. 1 in the Appendix \ref{sec:ATab}. The achieved storage times with these systems would enable a terrestrial test of our scheme. Combining a REID and an alkali gas QMem would result in $\Omega_->100$~THz, for example by combining a Pr and Rb QMem. In this case $\tau^\mathrm{reslim}$ is around a few fs which, \textcolor{black}{as has been stressed above,} recently been achieved in~\citep{Deschenes2016,Bergeron2019}. \textcolor{black}{Therefore, timing resolution requirements for terrestrial experiments in both domains are within today's technological reach.}

}

A great advantage of terrestrial experiments based on QMems would be that, in addition to the HOM experiment, a MZ experiment would also be possible as the vertical scale of the required interferometers is orders of magnitude smaller than for the previous photonic proposals that do not employ QMems~{\citep{Zych2012,pallister2017blueprint,Mohageg2021:deep}}
and thus is already within the experimental realm~\citep{Minar2008,Stockill2017, Yu2020, Lago-Rivera2021}.

Finally, we discuss the prospects of using simple delay lines for the demonstration of EDs due to gravity. Note that similar considerations can be found in \cite{pallister2017blueprint,Mohageg2021:deep}. From Fig.~\ref{fig:times}, one finds that, for an ambitious synchronization resolution of $\tau_\mathrm{res}\sim 1$~ps across GEO and Earth's surface (limited due to the effect of Earth's atmosphere), one needs a fibre delay line with a length of around 750~km (assuming a minimal attenuation factor $\sim 0.2\,$dB/m and a refractive index $\sim 1.47$). The associated loss of $\sim150$~dB would render any detection impossible. One would be bound to perform the experiment outside Earth's atmosphere to have the chance to obtain  $\tau_\mathrm{res}$ of the order of 1~fs. Then, a delay line of $\sim 6\,$km (corresponding to a loss of $\sim 40\%$ of the signal) would enable the detection of EDs due to gravity for frequency entangled photons at the two telecom wavelength 1550 nm and 1310 nm (maximal attenuation factor of $\sim 0.36\,$dB/m and refractive index $\sim 1.47$) and a setup of two satellites where one is located at a geosynchronus orbit and the other at 10,000 km above Earth's surface. Of course, the latter is not fulfilling our condition of stationarity with respect to the Earth's surface, however, for the duration of the experiment $\tau_d$ this is still a good approximation (see App. \ref{app:longredshift}). 

\section{Conclusions}

To conclude, we have investigated EDs due to gravity in MZ interferometry of two-frequency superposition states of single photons and HOM interferometry of frequency entangled two-photon states involving QMems. 
EDs due to gravity arise because of the geometry of the considered setup, where the arms of the interferometer correspond to different paths through spacetime. More specifically, in each arm, the photons pass through QMems that store the photonic correlations in localized memory excitations for an equivalent local storage time $\tau_s$. The QMems in different interferometer arms are located at different potential levels in the gravitational field of the earth. Redshift (or equivalently gravitational and relativistic time dilation) then implies that the states of the localized storage excitations evolve differently (as seen from a global coordinate-dependent perspective). This difference in evolution leads to entanglement between the spatial and frequency degrees of freedom of the restored photons in the readout process. We want to emphasize that EDs through localized excitation as an intermediate state is fundamentally different from EDs in previous proposals where only photons were considered and EDs arise due to their propagation through space (e.g. \citep{Zych2012,pallister2017blueprint,Mohageg2021:deep,Mieling2022}). Accordingly, EDs due to gravity of photons and QMems necessitates a different description as we have developed it for this article.

We have derived the detection statistics of photons behind the interfering beam splitter, i.e. the interference patterns, as a function of the storage time $\tau_s$. In the case of MZ interferometry of two-frequency superposition states of single photons, the interferogram shows two oscillations with $\tau_s$, a fast oscillation proportional to the average frequency and a beating proportional to the difference frequency. For HOM interferometry of frequency entangled two-photon states, only one oscillation appears, which is twice as fast as the beating oscillation 
of MZ interferometry. This feature is indicative of quantum-enhanced phase estimation, a phenomenon where the accumulated phase in an interference experiment scales with the number of particles involved. Quantum-enhanced phase estimation can also be achieved with particles in a NOON state \cite{lee2002quantum,Mieling2022}, wherein all photons co-propagate through an unknown but common path in the interferometer, usually of the MZ type.
In stark contrast, the HOM setup ensures that each photon in the pair propagates through a separate arm of the interferometer, as indicated by Eq. \eqref{EQ:A_IN_PS2b}.

We also gave a detailed analysis of the involved EDs due to gravity in terms of entanglement measures, the linear entropy and the negativity of the reduced density matrix of the spatial sub-state. In particular, the entanglement measures show oscillations with the difference frequency and the storage time that can be directly related to the oscillations of the interference patterns. Additionally, we compared our results to the similar case where optical delay lines instead of QMems are employed to realize long evolution times of photons at different gravitational potential levels. 

We have discussed the prospects of experimental realization of EDs due to gravity with implementations employing QMems and optical delay lines.  We have found that HOM interferometry of frequency entangled two-photon states has the clear advantage that the absence of high-frequency oscillations (that appear in the MZ case) implies that phase stability is not necessary and allows for a much smaller timing precision of storage and free-space propagation. 
Furthermore, we have found that the necessary storage times to perform the proposed tests of EDs due to gravity with QMems have already been realized in the lab. We conclude that chances are good that corresponding experiments
can be conducted with near future technology in satellite-based implementations and even in terrestrial implementations when QMems are employed. Such experimental verification of EDs due to gravity would constitute an experimental test of theoretical modeling combining a multi-particle effect predicted by the quantum theory of light and an effect predicted by general relativity.

It should be noted, however, that this effect already occurs in flat spacetime geometries, for example, Rindler space (corresponding to a uniformly accelerated frame that can be associated with a uniform gravitational field via the equivalence principle), since relativistic time dilation as a first order effect does not presuppose a non-trivial spacetime curvature. From a theoretical view point it would be interesting to investigate the contributions of curvature and higher order expansions of the metric tensor in a Post-Newtonian framework to the EDs and examine the experimental prerequisites  to detect these contributions like the required storage time and temporal resolution on the QMems. Also the investigation of non-locality on EDs due to gravity in Fermi normal coordinates appears as an interesting outlook for future considerations, as this would also address the question of how gravity influences quantum coherences and in particular genuine quantum properties like entanglement and indistinguishability of particles with a wave function of finite spatial extension, i.e. spatial wave packets, where the work on hand is rather concerned with the gravitational influences in the spectral and temporal domain. 

Our analysis is based on second quantization, which serves as a convenient tool for the analysis of interferometry of several indistinguishable particles and the development of theoretical methods to explicitly characterize multi-particle quantum phenomena under gravitational influences. Furthermore, our formal treatment explores a practical path that may be used to describe spacetime curvature related effects on quantum resources in a full quantum field theoretical description.

We have discussed in the main text and shown in the appendix that some entanglement correlations between different particles are robust against non-trivial EDs in the gravitational field of the earth. This is an important result for the design of entanglement based quantum technological applications employing non-local entanglement shared between possibly distant particles as, presently, one of the major challenges in entanglement based quantum technologies is the avoidance of decoherence of entanglement correlations.

\section*{Acknowledgements}

The authors thank Daniel Oi, Igor Pikovski, Albert Roura, Ernst M. Rasel, Rainer Dumke, Waldemar Herr, Arash Ahmadi, Erhan Sağlamyürek, Patrick Ledingham, Andreas W. Schell and David Edward Bruschi for useful discussions and comments about the topic of the article. R.B. acknowledges funding by the Deutsche Forschungsgemeinschaft (DFG, German Research Foundation) under Germany’s Excellence Strategy – EXC-2123 QuantumFrontiers – 390837967 and support by the Research Training Group 1620 “Models of Gravity”. D.R. acknowledges funding by the Federal Ministry of Education and Research of Germany in the project “Open6GHub” (grant number: 16KISK016) and support through the Deutsche Forschungsgemeinschaft (DFG, German Research Foundation) under Germany’s Excellence Strategy – EXC-2123 QuantumFrontiers – 390837967, the Research Training Group 1620 “Models of Gravity” and the TerraQ initiative from the Deutsche Forschungsgemeinschaft (DFG, German Research Foundation) – Project-ID 434617780 – SFB 1464. M.~G. and M.~K. acknowledge the support from DLR through funds provided by BMWi (OPTIMO, No.~50WM1958 and OPTIMO-II, No.~50WM2055). M.~G. further acknowledges funding from the European Union's Horizon 2020 research and innovation programme under the Marie Skłodowska-Curie grant agreement No.~894590. 

\bibliographystyle{quantum}
\bibliography{GIEDbib}

\appendix
\widetext

\section{Entangled clock states under EDs due to gravity \label{sec:A1}}

We consider two particles, which are both allowed to be situated at sites $U$ and $L$ and feature the same single-particle internal energy structure as in the single-particle example from the main text. The configuration space is now 16-dimensional, and we consider the two-particle system to be driven by the interaction-free two-particle Hamiltonian 
\begin{align}
    \hat{H}_2=\hat{H}_1 \otimes \mathbb{1}_{4 \times 4} +  \mathbb{1}_{4 \times 4} \otimes \hat{H}_1
\end{align} where $\mathbb{1}_{4 \times 4}$ is the identity operator on the four-dimensional single-particle Hilbert space. Analogously to the single-particle case, we can investigate the time evolution (from the perspective of an observer evolving w.r.t. to their proper time $\tau_o$) of some initial state $\ket{\psi(0)}$ by computing $\ket{\psi(\tau_o)}=e^{-i\hat{H}_2 \tau_o}\ket{\psi(0)}$. The resulting reduced $4 \times 4$ density matrix 
\begin{equation}
    \hat{\rho}_\mathrm{red}(\tau_o)=\sum_{i,j=1,2}\braket{\mu_i\mu_j|\hat{\rho}(\tau_o)|\mu_i\mu_j}
\end{equation} 
with $\hat{\rho}(\tau_o)=\ket{\psi(\tau_o)}\bra{\psi(\tau_o)}$ describes a two-qubit system in which the presence of entanglement can be unambiguously certified by positive values of the  \textit{negativity}
\begin{align}\label{EQ:Neg}
\mathcal{N}(\tau_o)=-\sum_{\lambda_{\mathrm{PT}}<0} \lambda_{\mathrm{PT}}(\tau_o), 
\end{align}
which is the negative sum of the negative eigenvalues $\lambda_{\mathrm{PT}}(\tau_o)$ of the partial transposed density matrix $\hat{\rho}_\mathrm{red}^\mathrm{PT}(\tau_o)=(\mathbb{1}_{4 \times 4} \otimes \hat{T})\hat{\rho}_\mathrm{red}(\tau_o)$, where $\hat{T}$ is the transposition operator \citep{HORODECKI19961}. If we consider an initial state 
\begin{align}
\ket{\psi(0)}=(\ket{\mu_1\mu_2}+\ket{\mu_2\mu_1}) \otimes (\ket{UL}+\ket{LU})/2    \label{EQ:AQ1}
\end{align}
in which both the internal and external DOFs of the two particles are entangled, the corresponding negativity
\begin{align}
    \mathcal{N}(\tau_o)=|\cos(\Delta_{\Theta^{}}\mu_-\tau_o)|/2
\end{align}
indicates that, after a time 
\begin{align}
\tau_{o,\mathrm{ent}}^{(2)}=\frac{\pi}{2\Delta_{\Theta^{}}\mu_-}\,,\label{EQ:tent2} 
\end{align}
the spatial entanglement between the two considered particles is depleted, where now the time to disentangle the spatial DOFs between the different particles is halved in comparison to the time needed to maximally entangle the internal and external DOF of the single-particle example from the main text, i.e. $\tau_{o,\mathrm{ent}}^{(2)}=\tau_{o,\mathrm{ent}}^{(1)}/2$, c.f. Eq (\ref{EQ:tent}) from the main text. Not only, the mutual spatial correlation between the particles gets lost. At the same time, entanglement correlations between the internal and external DOFs of the two-particle system are build up as can be seen by inspection of the linear entropy of the reduced spatial substate
\begin{align}
    \mathcal{S}(\tau_o)=1-\mathcal{P}(\tau_o)=\sin^2(\Delta_{\Theta}\mu_-\tau_o)/2,\label{EQ:A2Q}
\end{align}
where
\begin{align}
    \mathcal{P}(\tau_o)=\sum_{i,j=1,2}\braket{\sigma_i\sigma_j|\hat{\rho}^2_\mathrm{red}(\tau_o)|\sigma_i\sigma_j}
\end{align}
is the purity of the reduced spatial substate. In particular, any coherences of the spatial sub state decohered after the time $\tau_{o,\mathrm{ent}}^{(2)}$ since at this time, the spatial sub state $$\hat{\rho}_\mathrm{red}(\tau_{o,\mathrm{ent}}^{(2)})=(\ket{UL}\bra{UL}+\ket{LU}\bra{LU})/2$$ only features non-vanishing diagonal elements. It is interesting to notice that quantum states where the internal sub state is in an aligned Bell state and the positional sub state is in an anti-aligned Bell state or vice versa, i.e.
\begin{align}
    \ket{\bar{{\psi}}(0)}=(\ket{\mu_1\mu_2}+\ket{\mu_2\mu_1}) \otimes (\ket{UU}+\ket{LL})/2,    \label{EQ:AQ2}
    \\
    \ket{\tilde{\psi}(0)}=(\ket{\mu_1\mu_1}+\ket{\mu_2\mu_2}) \otimes (\ket{UL}+\ket{LU})/2,    \label{EQ:AQ3}
\end{align}
do not suffer from the EDs induced by $\hat{H}_2$ (since in this case one has $\mathcal{N}(\tau_o)=1/2$ for all $\tau_o$) which makes these states more attractive for quantum technological applications.

\section{Model for the quantum memories and imprinted phases  }\label{sec:QMemModel}

In this section, we present our model for the storage of the optical quantum states in QMems. Instead of focusing on a specific implementation, we assume general properties of the QMems. In particular, we assume that storage and read-out can be described by mode-swapper operations for each arm of the interferometer $\hat U_{\sigma}^{(\mathrm{map})} = \hat U_{\sigma 1}^{(\mathrm{map})} \hat U_{\sigma 2}^{(\mathrm{map})}$, where (s for storage and r for readout) (see for example \cite{choi2011coherent})
{
\begin{equation}
    \hat{U}_{\sigma i}^\mathrm{s/r} = \exp\left( \pi/2 \left(e^{i\phi_{\sigma i}^\mathrm{s/r}} a_{\sigma i} S_{\sigma i}^\dagger - e^{-i\phi_{\sigma i}^\mathrm{s/r}} a_{\sigma i}^\dagger S_{\sigma i}\right)\right)\,,
\end{equation}
}
and
\begin{equation}
    a_{\sigma i} = \int d\omega\, g_{\Omega_i\xi}(\omega)a_{\sigma\omega}
\end{equation}
is the anniliation operator for a photon at the spectral peak $g_{\Omega_i\xi}(\omega)$ and 
and $S_{\sigma i}^\dagger$ and $S_{\sigma i}$ are the creation and annihilation operators for excitations in the QMems. The phases $\phi_{\sigma i}^\mathrm{s/r}$ are imprinted in the storage/read-out process.

The photonic input state before the storage in the QMems is
\begin{align}
\ket{\psi_{\mathrm{MZ}}}_0=& N_{\mathrm{MZ}}[(\hat{a}_{U1}^\dagger-\hat{a}_{L1}^\dagger)+e^{i\varphi}(\hat{a}_{U2}^\dagger-\hat{a}_{L2}^\dagger)]  /\sqrt{2} \ket{0}  ,
\\
\ket{\psi_{\mathrm{HOM}}}_{0}=& N_{\mathrm{HOM}}[\hat{a}^\dagger_{U 1}\hat{a}^\dagger_{L 2}   +  e^{i\varphi}\hat{a}^\dagger_{U 2}\hat{a}^\dagger_{L 1}] \ket{0}\,.
\end{align}
We assume that the storage and release operations can be considered as instantaneous in comparison to the storage time. We find for the state in the QMem 
\begin{subequations}
\begin{align}
\ket{\psi_{\mathrm{MZ}}}_{\mathrm{QMem},0}=& N_{\mathrm{MZ}}[(e^{i\phi_{U1}^\mathrm{s}}\hat{S}_{U1}^\dagger-e^{i\phi_{L1}^\mathrm{s}}\hat{S}_{L1}^\dagger)+e^{i\varphi}(e^{i\phi_{U2}^\mathrm{s}}\hat{S}_{U2}^\dagger-e^{i\phi_{L2}^\mathrm{s}}\hat{S}_{L2}^\dagger)]  /\sqrt{2} \ket{0}  ,
\\
\ket{\psi_{\mathrm{HOM}}}_{\mathrm{QMem},0}=& N_{\mathrm{HOM}}[e^{i(\phi_{U1}^\mathrm{s}+\phi_{L2}^\mathrm{s})}\hat{S}^\dagger_{U 1}\hat{S}^\dagger_{L 2}   +  e^{i(\phi_{U2}^\mathrm{s}+\phi_{L1}^\mathrm{s})}e^{i\varphi} \hat{S}^\dagger_{U 2}\hat{S}^\dagger_{L 1}] \ket{0}\,.
\end{align}
\label{EQ:A_IN_PS_inQM}
\end{subequations}
When stored in the QMems, the time evolution of the excitations is defined by their respective energy $\tilde\nu_{\sigma 1}$ and $\tilde\nu_{\sigma 2}$ and the local proper storage times $\tau_{\sigma s}$. We introduce a parameter $\chi$ that is 1 if the  internal phase evolution is imprinted on the output state of the QMem and 0 otherwise, and we find 
{ 
\begin{subequations}
\begin{align}
\ket{\psi_{\mathrm{MZ}}}_{\mathrm{QMem},\mathrm{f}}=& N_{\mathrm{MZ}}[(\hat{S}_{U1}^\dagger e^{i(\chi\tilde\nu_{U1} \tau_{U\mathrm{s}} + \phi_{U1}^\mathrm{s})} - \hat{S}_{L1}^\dagger e^{i(\chi\tilde\nu_{L1} \tau_{L\mathrm{s}} + \phi_{L1}^\mathrm{s})}) \nonumber \\
& + e^{i\varphi}(\hat{S}_{U2}^\dagger e^{i(\chi\tilde\nu_{U2}  \tau_{U\mathrm{s}} + \phi_{U2}^\mathrm{s})} - \hat{S}_{L2}^\dagger e^{i(\chi\tilde\nu_{L2}  \tau_{L\mathrm{s}}+\phi_{L2}^\mathrm{s}}))]  /\sqrt{2} \ket{0},
\\
\ket{\psi_{\mathrm{HOM}}}_{\mathrm{QMem},\mathrm{f}}=& N_{\mathrm{HOM}}[S^\dagger_{U 1}S^\dagger_{L 2}e^{i(\chi\tilde\nu_{U1}\tau_{U\mathrm{s}} + \chi\tilde\nu_{L2} \tau_{L\mathrm{s}} + \phi_{U1}^\mathrm{s} + \phi_{L2}^\mathrm{s})}   +  e^{i\varphi} S^\dagger_{U 2}S^\dagger_{L 1} e^{i(\chi\tilde\nu_{U2}\tau_{U\mathrm{s}} + \chi\tilde\nu_{L1}\tau_{L\mathrm{s}} + \phi_{U2}^\mathrm{s} + \phi_{L1}^\mathrm{s})}] \ket{0}\,.
\end{align}
\label{EQ:A_IN_PS_inQM_storage}
\end{subequations}
}

Up to a global phase, the resulting state after the storage becomes
\begin{subequations}
\label{EQ:Apfbs}
\begin{align}
\ket{\psi_{\mathrm{MZ}}'}=& \sum_\sigma \int d\omega\, \psi_{\sigma}'(\omega)\hat{a}^\dagger_{\sigma \omega} \ket{0},
\\
\ket{\psi_{\mathrm{HOM}}'}=&\int d\omega_1 d\omega_2\, \psi'(\omega_1,\omega_2)a^\dagger_{U \omega_1}a^\dagger_{L \omega_2}\ket{0}\,,
\end{align}
\label{EQ:A_IN_PS2_app}
\end{subequations}
{
\begin{subequations}
\begin{align}
\psi_{\sigma}'(\omega)=&(-1)^{\tilde{n}(\sigma)} N_{\mathrm{MZ}}[g_{\Omega_1\xi}(\omega) e^{i(\chi\tilde\nu_{\sigma 1}\tau_{\sigma\mathrm{s}} + \phi_{\sigma 1}^\mathrm{s} - \phi_{\sigma 1}^\mathrm{r} )} + e^{i\varphi}g_{\Omega_2\xi}(\omega)e^{i(\chi\tilde\nu_{\sigma 2}\tau_{\sigma\mathrm{s}} + \phi_{\sigma 2}^\mathrm{s} - \phi_{\sigma 2}^\mathrm{r} )} ]/\sqrt{2}
\\
\psi'(\omega_1,\omega_2)=& N_{\mathrm{HOM}}[g_{\Omega_1\xi}(\omega_1)g_{\Omega_2\xi}(\omega_2)+e^{i\varphi'_{\mathrm{HOM}}}g_{\Omega_2\xi}(\omega_1)g_{\Omega_1\xi}(\omega_2)]\,,\label{EQ:APsiHOM_app}
\end{align}
\label{EQ:APsiApp}
\end{subequations}
where $\tilde{n}(U)=0$, $\tilde{n}(L)=1$ and 
\begin{align}
    \varphi'_{\mathrm{HOM}}=& \varphi + \chi(\tilde\nu_{U-}\tau_{U\mathrm{s}} - \tilde\nu_{L-}\tau_{L\mathrm{s}}) + ((\phi_{U 2}^s - \phi_{U 1}^s) - (\phi_{L 2}^s - \phi_{L 1}^s)) - ((\phi_{U 2}^r - \phi_{U 1}^r) - (\phi_{L 2}^r - \phi_{L 1}^r)) \,
\end{align}}
and $\tilde\nu_{\sigma -}= \tilde\nu_{\sigma 2}-\tilde\nu_{\sigma 1}$.

We assume that an independent external phase reference (e.g. a laser serving as a local local oscillator) is provided for $U$ and $L$ each which operate at the frequencies $\tilde\Omega^{(r)}_{\sigma i}$, for $i=1,2$ and $\sigma =U,L$, which are measured in the common time frame of the photon source. The additional phase difference which arises due to the usage of independent local phase references is $\phi_{\sigma i}^\mathrm{s} -\phi_{\sigma i}^\mathrm{r} = \tilde\Omega^{(r)}_{\sigma i} \tau_{\sigma s}$. In this case, we find
{
\begin{align}
\psi_{\sigma}'(\omega)=& (-1)^{\tilde{n}(\sigma)} N_{\mathrm{MZ}}[g_{\Omega_1\xi}(\omega) e^{i(\chi\tilde\nu_{\sigma 1} + \tilde\Omega^{(r)}_{\sigma 1}) \tau_{\sigma\mathrm{s}}} + e^{i\varphi}g_{\Omega_2\xi}(\omega)e^{i(\chi\tilde\nu_{\sigma 2} + \tilde\Omega^{(r)}_{\sigma 2}) \tau_\mathrm{\sigma s}} ]/\sqrt{2}
\end{align}
and
\begin{align}
    \varphi'_{\mathrm{HOM}}=&  (\chi\tilde\nu_{U-} + \tilde\Omega^{(r)}_{U-})\tau_{U\mathrm{s}} - (\chi\tilde\nu_{L-} + \tilde\Omega^{(r)}_{L-})\tau_{L\mathrm{s}} + \varphi\,,
\end{align}
where $\tilde\Omega^{(r)}_{\sigma -} = \tilde\Omega^{(r)}_{\sigma 2}-\tilde\Omega^{(r)}_{\sigma 1}$. Taking into account that $\tilde\nu_{\sigma i} + \tilde\Omega^{(r)}_{\sigma i} = \tilde\Omega_{\sigma i}$, we find
\begin{align}\label{eq:psiprimeMZ}
\psi_{\sigma}'(\omega)=& (-1)^{\tilde{n}(\sigma)} N_{\mathrm{MZ}}[g_{\Omega_1\xi}(\omega) e^{i(\tilde\Omega_{\sigma 1}+(\chi-1)\tilde\nu_{\sigma 1})\tau_{\sigma\mathrm{s}}} + e^{i\varphi} g_{\Omega_2\xi}(\omega)e^{i(\tilde\Omega_{\sigma 2} + (\chi-1)\tilde\nu_{\sigma 2}) \tau_\mathrm{\sigma s}} ]/\sqrt{2}
\end{align}
and
\begin{align}\label{eq:varphiprimeHOM}
    \varphi'_{\mathrm{HOM}}=&\,((\chi-1)\tilde\nu_{U-} + \tilde\Omega_{U-})\tau_{U\mathrm{s}} - ((\chi-1)\tilde\nu_{L-} + \tilde\Omega_{L-})\tau_{L\mathrm{s}} + \varphi \\
\nonumber =&\,  \Omega_{-}(\tau_{U\mathrm{s}}/\Theta_U - \tau_{L\mathrm{s}}/\Theta_L) + (\chi-1)(\tilde\nu_{U-}\tau_{U\mathrm{s}} - \tilde\nu_{L-} \tau_{L\mathrm{s}}) + \varphi \,.
\end{align}
{with}  $\tilde\Omega_{\sigma \pm} = \tilde\Omega_{\sigma 2}\pm\tilde\Omega_{\sigma 1}$, $\tilde\nu_{\sigma \pm} = \tilde\nu_{\sigma 2}\pm\tilde\nu_{\sigma 1}$ and $\Omega_{\pm} = \Omega_{2}\pm\Omega_{1}${, where the relation between the $i$-th spectral peak frequency $\tilde\Omega_{\sigma i}$ as perceived from the local rest frame of the QMems at $\sigma=U,L$ to the respective frequency peak $\Omega_{i}$ as perceived from the rest frame of the photon source is given by $\tilde\Omega_{\sigma i}=\Omega_{i}/\Theta_\sigma$}.

}

{ Next, we derive the MZ- and HOM-interference patterns. First, we apply a $50:50$ beam splitter to Eq. (\ref{EQ:Apfbs}) which transforms the photonic creation operators as $\hat{a}_{U\omega}\rightarrow (\hat{a}_{\oplus\omega}-\hat{a}_{\ominus\omega})/\sqrt{2}$ and $\hat{a}_{L\omega}\rightarrow (\hat{a}_{\oplus\omega}+\hat{a}_{\ominus\omega})/\sqrt{2}$. After the beam splitter, we have
\begin{subequations}
\begin{align}
\ket{\psi_{\mathrm{MZ}}''}=& \int d\omega\,( \hat{a}^\dagger_{\oplus\omega}(\psi_{U}'(\omega) +
\psi_{L}'(\omega)) - \hat{a}^\dagger_{\ominus\omega}(\psi_{U}'(\omega) - \psi_{L}'(\omega)))/\sqrt{2}\ket{0},
\\
\ket{\psi_{\mathrm{HOM}}''}=&\int d\omega_1 d\omega_2\, \psi'(\omega_1,\omega_2) (\hat{a}_{\oplus\omega_1}^\dagger-\hat{a}_{\ominus\omega_1}^\dagger)(\hat{a}_{\oplus\omega_2}^\dagger+\hat{a}_{\ominus\omega_2}^\dagger)/2\ket{0},
\end{align}
\label{EQ:A_IN_PS2_app2}
\end{subequations}
which both can be represented in the form of (\ref{EQ:PS}), that is 
\begin{subequations}
\begin{align}
\ket{\psi_{\mathrm{MZ}}''}=&\sum_\sigma\int d\omega\, \psi^\mathrm{f}_{\sigma}(\omega)\hat{a}^\dagger_{\sigma \omega}\ket{0},    
\\
\ket{\psi_{\mathrm{HOM}}''}=&\frac{1}{\sqrt{2}}\sum_{\sigma_1,\sigma_2}\int d\omega_1 d\omega_2\, \psi^\mathrm{f}_{\sigma_1\sigma_2}(\omega_1,\omega_2)\hat{a}^\dagger_{\sigma_1 \omega_1}\hat{a}^\dagger_{\sigma_2 \omega_2}\ket{0},    
\end{align}
\end{subequations}
where one has 
\begin{subequations}
\begin{align}
\psi^\mathrm{f}_{\sigma}(\omega)=&\frac{1}{\sqrt{2}} [\delta_{\sigma \oplus}(\psi_{U}'(\omega) +\psi_{L}'(\omega)) - \delta_{\sigma \ominus}(\psi_{U}'(\omega) - \psi_{L}'(\omega))],\label{EQ:APF1}
\\
\nonumber \psi^\mathrm{f}_{\sigma_1\sigma_2}(\omega_1,\omega_2) &=\, \frac{1}{2\sqrt{2}}\Big[\psi'(\omega_1,\omega_2)(\delta_{\sigma_1 \oplus}\delta_{\sigma_2 \oplus}-\delta_{\sigma_1 \ominus}\delta_{\sigma_2 \ominus}+\delta_{\sigma_1 \oplus}\delta_{\sigma_2 \ominus}-\delta_{\sigma_1 \ominus}\delta_{\sigma_2 \oplus}) + ((\sigma_1,\omega_1)\longleftrightarrow (\sigma_2,\omega_2)) \Big] \label{EQ:APF2}\\
\nonumber  &=\, \frac{1}{2\sqrt{2}}\Big[(\psi'(\omega_1,\omega_2) + \psi'(\omega_2,\omega_1))(\delta_{\sigma_1 \oplus}\delta_{\sigma_2 \oplus}-\delta_{\sigma_1 \ominus}\delta_{\sigma_2 \ominus}) \\
& \quad + (\psi'(\omega_1,\omega_2) - \psi'(\omega_2,\omega_1))(\delta_{\sigma_1 \oplus}\delta_{\sigma_2 \ominus}-\delta_{\sigma_1 \ominus}\delta_{\sigma_2 \oplus}) \Big] \\
\nonumber  &= \, \frac{N_{\mathrm{HOM}}}{2\sqrt{2}}\Big[(1+e^{i\varphi'_{\mathrm{HOM}}})(g_{\Omega_1\xi}(\omega_1)g_{\Omega_2\xi}(\omega_2)+g_{\Omega_2\xi}(\omega_1)g_{\Omega_1\xi}(\omega_2))(\delta_{\sigma_1 \oplus}\delta_{\sigma_2 \oplus}-\delta_{\sigma_1 \ominus}\delta_{\sigma_2 \ominus}) \\
\nonumber   & \quad + (1-e^{i\varphi'_{\mathrm{HOM}}})(g_{\Omega_1\xi}(\omega_1)g_{\Omega_2\xi}(\omega_2)-g_{\Omega_2\xi}(\omega_1)g_{\Omega_1\xi}(\omega_2))(\delta_{\sigma_1 \oplus}\delta_{\sigma_2 \ominus}-\delta_{\sigma_1 \ominus}\delta_{\sigma_2 \oplus}) \Big]   \\
\nonumber &= \, \frac{N_{\mathrm{HOM}}}{\sqrt{2}}e^{i\varphi'_{\mathrm{HOM}}/2}\Big[\cos\left(\varphi'_{\mathrm{HOM}}/2\right)(g_{\Omega_1\xi}(\omega_1)g_{\Omega_2\xi}(\omega_2)+g_{\Omega_2\xi}(\omega_1)g_{\Omega_1\xi}(\omega_2))(\delta_{\sigma_1 \oplus}\delta_{\sigma_2 \oplus}-\delta_{\sigma_1 \ominus}\delta_{\sigma_2 \ominus}) \\
\nonumber   & \quad - i \sin\left(\varphi'_{\mathrm{HOM}}/2\right)(g_{\Omega_1\xi}(\omega_1)g_{\Omega_2\xi}(\omega_2)-g_{\Omega_2\xi}(\omega_1)g_{\Omega_1\xi}(\omega_2))(\delta_{\sigma_1 \oplus}\delta_{\sigma_2 \ominus}-\delta_{\sigma_1 \ominus}\delta_{\sigma_2 \oplus}) \Big] \,,
\end{align}
\label{EQ:A_MZHOM}
\end{subequations}
where $\sigma=\oplus,\ominus$. 
From Eqs. \eqref{eq:diagentriesredrho} from the main text, i.e.
{
\begin{subequations}
\label{eq:diagentriesredrhoApp}
\begin{align}\label{eq:PsigmaApp}
    P_{\sigma'}^{\mathrm{MZ}}&=\int d\omega\,\braket{\hat{a}^\dagger_{\sigma'\omega}\hat{a}_{\sigma'\omega}}=\rho_{\sigma'\sigma'}\quad\mathrm{and}\\
    \label{eq:Psigma1sigma2App}
   \nonumber  P_{\sigma_1'\sigma_2'}^{\mathrm{HOM}}&=\int d\omega_1 d\omega_2\,  \braket{\hat{a}^\dagger_{\sigma_1'\omega_1}\hat{a}^\dagger_{\sigma_2'\omega_2}\hat{a}_{\sigma_2'\omega_2}\hat{a}_{\sigma_1'\omega_1}}{/2}\\
    & =\rho_{\sigma_1'\sigma_2'\sigma_1'\sigma_2'} \,,
\end{align}
\end{subequations}
}
with $\sigma=\oplus,\ominus$, we obtain the probabilities of the various detection events and combine them according to Eq. \eqref{eq:PcMZ} and \eqref{eq:PcHOM}, that is,
 and $P_c^{\mathrm{MZ}}=P_\ominus^{\mathrm{MZ}} - P_\oplus^{\mathrm{MZ}}$ as well as $P_c^{\mathrm{HOM}}=(P_{\oplus\oplus}^{\mathrm{HOM}}+P_{\ominus\ominus}^{\mathrm{HOM}})-(P_{\oplus\ominus}^{\mathrm{HOM}}+P_{\ominus\oplus}^{\mathrm{HOM}})$.

Inserting (\ref{eq:psiprimeMZ}) into Eq. (\ref{EQ:APsi}) with \eqref{eq:varphiprimeHOM} and taking the corresponding diagonal elements of the resulting reduced density matrices yields
\begin{align}
\nonumber P_c^{\mathrm{MZ}}=&\, -\int d\omega (\psi_{U}'(\omega)(\psi_{L}'(\omega))^* + \psi_{L}'(\omega)(\psi_{U}'(\omega))^*)\\
 =&\, \sum_i \cos((\tilde\Omega_{Ui} + (\chi-1)\tilde\nu_{Ui})\tau_{U\mathrm{s}}-(\tilde\Omega_{Li} + (\chi-1)\tilde\nu_{Li})\tau_{L\mathrm{s}})\,N_{\mathrm{MZ}}^2\int d\omega\, g_{\Omega_i\xi}(\omega)^2 \\
\nonumber &\,   +\Big(\cos((\tilde\Omega_{U1} + (\chi-1)\tilde\nu_{U1})\tau_{U\mathrm{s}}-(\tilde\Omega_{L2} + (\chi-1)\tilde\nu_{L2}) \tau_{L\mathrm{s}} - \varphi) \\
\nonumber &\quad +  \cos((\tilde\Omega_{L1} + (\chi-1)\tilde\nu_{L1})\tau_{L\mathrm{s}}-(\tilde\Omega_{U2} + (\chi-1)\tilde\nu_{U2})\tau_{U\mathrm{s}} - \varphi)\Big) N_{\mathrm{MZ}}^2\int d\omega\, g_{\Omega_1\xi}(\omega)g_{\Omega_2\xi}(\omega) \\
\nonumber  =&\, 2N_{\mathrm{MZ}}^2\cos((\tilde\Omega_{U+} + (\chi-1)\tilde\nu_{U+})\tau_{U\mathrm{s}}/2 - (\tilde\Omega_{L+} + (\chi-1)\tilde\nu_{L+})\tau_{L\mathrm{s}}/2)\times\\
\nonumber &\, \times\cos((\tilde\Omega_{U-} + (\chi-1)\tilde\nu_{U-})\tau_{U\mathrm{s}}/2 - (\tilde\Omega_{L-} + (\chi-1)\tilde\nu_{L-})\tau_{L\mathrm{s}}/2)\\
\nonumber &\,   +2N_{\mathrm{MZ}}^2e^{-\Omega_-^2/8\xi^2} \cos((\tilde\Omega_{U+} + (\chi-1)\tilde\nu_{U+})\tau_{U\mathrm{s}}/2 - (\tilde\Omega_{L+} + (\chi-1)\tilde\nu_{L+})\tau_{L\mathrm{s}}/2)\times\\
\nonumber &\, \times\cos((\tilde\Omega_{U-} + (\chi-1)\tilde\nu_{U-})\tau_{U\mathrm{s}}/2 + (\tilde\Omega_{L-} + (\chi-1)\tilde\nu_{L-})\tau_{L\mathrm{s}}/2 + \varphi) \\
\nonumber  =&\, 2N_{\mathrm{MZ}}^2\cos(\Omega_{+}\tau_{U\mathrm{s}}/2\Theta_U - \Omega_{+}\tau_{L\mathrm{s}}/2\Theta_L +  (\chi-1)(\tilde\nu_{U+}\tau_{U\mathrm{s}} - \tilde\nu_{L+}\tau_{L\mathrm{s}})/2)\times\\
\nonumber &\,\times \cos(\Omega_{-}\tau_{U\mathrm{s}}/2\Theta_U - \Omega_{-}\tau_{L\mathrm{s}}/2\Theta_L + (\chi-1)(\tilde\nu_{U-}\tau_{U\mathrm{s}} - \tilde\nu_{L-}\tau_{L\mathrm{s}})/2)\\
\nonumber &\,  +2N_{\mathrm{MZ}}^2 e^{-\Omega_-^2/8\xi^2}\cos(\Omega_{+}\tau_{U\mathrm{s}}/2\Theta_U - \Omega_{+}\tau_{L\mathrm{s}}/2\Theta_L +  (\chi-1)(\tilde\nu_{U+}\tau_{U\mathrm{s}} - \tilde\nu_{L+}\tau_{L\mathrm{s}})/2) \times\\
\nonumber &\,\times \cos(\Omega_{-}\tau_{U\mathrm{s}}/2\Theta_U + \Omega_{-}\tau_{L\mathrm{s}}/2\Theta_L +  (\chi-1)(\tilde\nu_{U-}\tau_{U\mathrm{s}} + \tilde\nu_{L-}\tau_{L\mathrm{s}})/2 + \varphi) 
\,,
\end{align}
where $\Omega_{\pm}=\Omega_{2}\pm\Omega_{1}$. For HOM interference, we find
\begin{align}
\nonumber \rho_{\sigma_1\sigma_2\bar{\sigma}_1\bar{\sigma}_2} & =\, \int d\omega_1 d\omega_2\, \psi^\mathrm{f}_{ \sigma_1\sigma_2}(\omega_1,\omega_2) \psi^\mathrm{f}_{ \bar{\sigma}_1\bar{\sigma}_2}(\omega_1,\omega_2)^* \\
&= \frac{1}{8}  \int d\omega_1 d\omega_2\, \left|\psi'(\omega_1,\omega_2) + \psi'(\omega_2,\omega_1)\right|^2 (\delta_{\sigma_1 \oplus}\delta_{\sigma_2 \oplus}-\delta_{\sigma_1 \ominus}\delta_{\sigma_2 \ominus})(\delta_{\bar\sigma_1 \oplus}\delta_{\bar\sigma_2 \oplus}-\delta_{\bar\sigma_1 \ominus}\delta_{\bar\sigma_2 \ominus})  \\
\nonumber & \quad + \frac{1}{8}  \int d\omega_1 d\omega_2\, \left|\psi'(\omega_1,\omega_2) - \psi'(\omega_2,\omega_1)\right|^2 (\delta_{\sigma_1 \oplus}\delta_{\sigma_2 \ominus}-\delta_{\sigma_1 \ominus}\delta_{\sigma_2 \oplus})(\delta_{\bar\sigma_1 \oplus}\delta_{\bar\sigma_2 \ominus}-\delta_{\bar\sigma_1 \ominus}\delta_{\bar\sigma_2 \oplus}) \,.
\end{align}
Accordingly, we obtain 
\begin{align}
\nonumber P_c^{\mathrm{HOM}}=&  \, (P_{\oplus\oplus}^{\mathrm{HOM}}+P_{\ominus\ominus}^{\mathrm{HOM}})-(P_{\oplus\ominus}^{\mathrm{HOM}}+P_{\ominus\oplus}^{\mathrm{HOM}}) \\
\nonumber =&\, \frac{1}{4} \int d\omega_1 d\omega_2\,  \left( \left|\psi'(\omega_1,\omega_2) + \psi'(\omega_2,\omega_1)\right|^2 - \left|\psi'(\omega_1,\omega_2) - \psi'(\omega_2,\omega_1)\right|^2\right) \\
=&\, \frac{1}{2} \int d\omega_1 d\omega_2\,  \left( \psi'(\omega_1,\omega_2)\psi'(\omega_2,\omega_1)^* + \psi'(\omega_1,\omega_2)^*\psi'(\omega_2,\omega_1)\right)  \\
\nonumber =&\,  \Re\left\{\int d\omega_1 d\omega_2\,   \psi'(\omega_1,\omega_2)\psi'(\omega_2,\omega_1)^* \right\} \\
\nonumber =& \,  N_{\mathrm{HOM}}^2 \Re\Bigg\{\int d\omega_1 d\omega_2\,   [g_{\Omega_1\xi}(\omega_1)g_{\Omega_2\xi}(\omega_2)+e^{i\varphi'_{\mathrm{HOM}}}g_{\Omega_2\xi}(\omega_1)g_{\Omega_1\xi}(\omega_2)] \times \\
\nonumber & \quad \times [g_{\Omega_1\xi}(\omega_2)g_{\Omega_2\xi}(\omega_1)+e^{-i\varphi'_{\mathrm{HOM}}}g_{\Omega_2\xi}(\omega_2)g_{\Omega_1\xi}(\omega_1)] \Bigg\} \\
\nonumber =& \,  2N_{\mathrm{HOM}}^2 \left[ \cos\left(\varphi'_{\mathrm{HOM}}\right)  + \left(\int d\omega \,g_{\Omega_1\xi}(\omega)g_{\Omega_2\xi}(\omega) \right)^2  \right] \\
\nonumber =& \, 2N_{\mathrm{HOM}}^2 \left[ \cos\left(\varphi'_{\mathrm{HOM}}\right)  + e^{-\Omega_-^2/4\xi^2}  \right] \\
\nonumber =& \, \frac{ \cos\left(\varphi'_{\mathrm{HOM}}\right)  + e^{-\Omega_-^2/4\xi^2} }{1 + \cos\left(\varphi\right) e^{-\Omega_-^2/4\xi^2}}  \,.
\end{align}
In the limit $\Omega_-\gg \xi$, the interference patterns simplify to
\begin{subequations}
\begin{align}
   P_c^{\mathrm{MZ}}=&\, \cos(\Omega_{+}(\tau_{U\mathrm{s}}/\Theta_U - \tau_{L\mathrm{s}}/\Theta_L)/2 + (\chi-1)(\tilde\nu_{U+}\tau_{U\mathrm{s}} - \tilde\nu_{L+}\tau_{L\mathrm{s}})/2)\times\\
   \nonumber &\,\times \cos(\Omega_{-}(\tau_{U\mathrm{s}}/\Theta_U - \tau_{L\mathrm{s}}/\Theta_L)/2 + (\chi-1)(\tilde\nu_{U-}\tau_{U\mathrm{s}} - \tilde\nu_{L-}\tau_{L\mathrm{s}})/2)\\
   P_c^{\mathrm{HOM}}=&\, \cos\left(\varphi'_{\mathrm{HOM}}\right) = \cos\left(\Omega_{-}(\tau_{U\mathrm{s}}/\Theta_U - \tau_{L\mathrm{s}}/\Theta_L) + (\chi-1)(\tilde\nu_{U-}\tau_{U\mathrm{s}} - \tilde\nu_{L-}\tau_{L\mathrm{s}}) + \varphi \right)\,,
\end{align}
\end{subequations}
and if $\chi=1$
\begin{subequations}
\begin{align}
   P_c^{\mathrm{MZ}}=&\, \cos(\Omega_{+}(\tau_{U\mathrm{s}}/\Theta_U - \tau_{L\mathrm{s}}/\Theta_L)/2)\cos(\Omega_{-}(\tau_{U\mathrm{s}}/\Theta_U - \tau_{L\mathrm{s}}/\Theta_L)/2)\\
   P_c^{\mathrm{HOM}}=&\, \cos\left(\varphi'_{\mathrm{HOM}}\right) = \cos\left(\Omega_{-}(\tau_{U\mathrm{s}}/\Theta_U - \tau_{L\mathrm{s}}/\Theta_L)  + \varphi \right)\,.
\end{align}
\end{subequations}

}


\section{Derivation of the MZ- and HOM-interference patterns for delay lines}\label{sec:dervPat}

{
Here we derive the MZ- and HOM-interference patterns for the case that optical delays are applied in the transmission paths from Fig. \ref{fig:HOM} from the main text instead of QMems. We start from the initial states
\begin{subequations}
\begin{align}
\ket{\psi_{\mathrm{MZ}}}=&\int d\omega\, \psi(\omega)a^\dagger_{A \omega}\ket{0},
\\
\ket{\psi_{\mathrm{HOM}}}=&\int d\omega_1 d\omega_2\, \psi(\omega_1,\omega_2)a^\dagger_{U \omega_1}a^\dagger_{L \omega_2}\ket{0},
\end{align}
\label{EQ:A_IN_PS}
\end{subequations}
which are injected into the respective interferometer in Fig. \ref{fig:HOM} from the main text. In case of the MZ-interferometer, we have a first beam splitter which transforms the photonic creation operators as $\hat{a}_{A\omega}\rightarrow (\hat{a}_{U\omega}-\hat{a}_{L\omega})/\sqrt{2}$ and $\hat{a}_{B\omega}\rightarrow (\hat{a}_{U\omega}+\hat{a}_{L\omega})/\sqrt{2}$. 
Both, the MZ- and HOM-interferometer employ optical delays, which transform the photonic creation operators as 
 $\hat{a}_{\sigma\omega}\rightarrow \hat{a}_{\sigma\omega}e^{i\omega \tau_\mathrm{d}/\Theta_\sigma}$ with $\sigma=U,L$ (c.f. Fig. \ref{fig:HOM} from the main text). Then, a 50:50 beam splitter is applied, which transforms the photonic creation operators as $\hat{a}_{U\omega}\rightarrow (\hat{a}_{\oplus\omega}-\hat{a}_{\ominus\omega})/\sqrt{2}$ and $\hat{a}_{L\omega}\rightarrow (\hat{a}_{\oplus\omega}+\hat{a}_{\ominus\omega})/\sqrt{2}$. 
 
 Passing the respective sequence of beam splitters and optical delays, the quantum states (\ref{EQ:A_IN_PS}) transform into 
\begin{subequations}
\begin{align}
\ket{\psi'_{\mathrm{MZ}}}=&\int d\omega\, \psi(\omega)\left[a^\dagger_{\oplus \omega}(e^{i\omega \tau_\mathrm{d}/\Theta_U}-e^{i\omega \tau_\mathrm{d}/\Theta_L})-a^\dagger_{\ominus \omega}(e^{i\omega \tau_\mathrm{d}/\Theta_U}+e^{i\omega \tau_\mathrm{d}/\Theta_L})\right]/2\ket{0}\label{EQ:APMZ}
\\
\ket{\psi'_{\mathrm{HOM}}}
=&\int d\omega_1 d\omega_2\, \psi(\omega_1,\omega_2) e^{i\omega_1 \tau_\mathrm{d}/\Theta_U}e^{i\omega_2 \tau_\mathrm{d}/\Theta_L}
(\hat{a}_{\oplus\omega_1}^\dagger-\hat{a}_{\ominus\omega_1}^\dagger)(\hat{a}_{\oplus\omega_2}^\dagger+\hat{a}_{\ominus\omega_2}^\dagger)/2\ket{0},
\end{align}
\end{subequations}
which both can be represented in the form of (\ref{EQ:PS}), where one has 
\begin{subequations}
\begin{align}
\psi_{\sigma}(\omega)=&\,\frac{1}{2}\psi(\omega)[\delta_{\sigma \oplus}(e^{i\omega \tau_\mathrm{d}/\Theta_U}-e^{i\omega \tau_\mathrm{d}/\Theta_L}) - \delta_{\sigma \ominus}(e^{i\omega \tau_\mathrm{d}/\Theta_U}+e^{i\omega \tau_\mathrm{d}/\Theta_L})],
\\
\nonumber \psi_{\sigma_1\sigma_2}(\omega_1,\omega_2)=& \frac{1}{2\sqrt{2}}\left[\psi(\omega_1,\omega_2)  e^{i\omega_1 \tau_\mathrm{d}/\Theta_U}e^{i\omega_2 \tau_\mathrm{d}/\Theta_L}(\delta_{\sigma_1 \oplus}-\delta_{\sigma_1 \ominus})(\delta_{\sigma_2 \oplus}+\delta_{\sigma_2 \ominus}) + ((\sigma_1,\omega_1)\longleftrightarrow (\sigma_2,\omega_2)) \right]\\
 =&\,  \frac{1}{2\sqrt{2}}\Bigg[\left(\psi(\omega_1,\omega_2)  e^{i\omega_1 \tau_\mathrm{d}/\Theta_U}e^{i\omega_2 \tau_\mathrm{d}/\Theta_L}+\psi(\omega_2,\omega_1)  e^{i\omega_2 \tau_\mathrm{d}/\Theta_U}e^{i\omega_1 \tau_\mathrm{d}/\Theta_L}\right)(\delta_{\sigma_1 \oplus}\delta_{\sigma_2 \oplus}+\delta_{\sigma_1 \ominus}\delta_{\sigma_2 \ominus}) \\
\nonumber &\, \left(\psi(\omega_1,\omega_2)  e^{i\omega_1 \tau_\mathrm{d}/\Theta_U}e^{i\omega_2 \tau_\mathrm{d}/\Theta_L}-\psi(\omega_2,\omega_1)  e^{i\omega_2 \tau_\mathrm{d}/\Theta_U}e^{i\omega_1 \tau_\mathrm{d}/\Theta_L}\right)(\delta_{\sigma_1 \oplus}\delta_{\sigma_2 \ominus}-\delta_{\sigma_1 \ominus}\delta_{\sigma_2 \oplus})  \Bigg],
\end{align}
\label{EQ:A_MZHOM_2}
\end{subequations}
where $\sigma=\oplus,\ominus$. Inserting (\ref{EQ:A_MZHOM_2}) into Eq. (\ref{EQ:rhomat}) from the main text and taking the corresponding diagonal elements of the resulting reduced density matrices yields
\begin{subequations}
\label{EQ:PC}
\begin{align}
P_c^{\mathrm{MZ}}=&\int d\omega |\psi(\omega)|^2\cos(\omega\Delta_{\Theta^{-1}} \tau_\mathrm{d}),
\\
P_c^{\mathrm{HOM}}=&\Re\left\{\int d\omega_1 d\omega_2 \psi(\omega_1,\omega_2)\psi^*(\omega_2,\omega_1)e^{i\Delta\omega\Delta_{\Theta^{-1}} \tau_\mathrm{d}}\right\},
\end{align}
\end{subequations}
where $\Delta\omega=\omega_2-\omega_1$ and $\Delta_{\Theta^{-1}}=1/\Theta_L - 1/\Theta_U$.

Inserting the 
spectra (\ref{EQ:SpecMZ}) and (\ref{EQ:SpecHOM}) from the main text into (\ref{EQ:PC}) results in
\begin{subequations}\label{eq:PcXfull}
\begin{align}
P_c^{\mathrm{X}}=&2N_{\mathrm{X}}^2\left[ R_{\mathrm{X}}+S_{\mathrm{X}}\right],\label{EQ:IP}
\\
R_{\mathrm{MZ}}=&e^{-\frac{1}{2}(\Delta_{\Theta^{-1}} \tau_\mathrm{d}\xi)^2}\cos(\Omega_+\Delta_{\Theta^{-1}} \tau_\mathrm{d}/2)\cos(\Omega_-\Delta_{\Theta^{-1}} \tau_\mathrm{d}/2)
\\
R_{\mathrm{HOM}}=&e^{-(\Delta_{\Theta^{-1}} \tau_\mathrm{d}\xi)^2}\cos(\Omega_-\Delta_{\Theta^{-1}} \tau_\mathrm{d}-\varphi)
\\
S_{\mathrm{MZ}}=&e^{-\Omega_-^2/(8\xi^2)}e^{-\frac{1}{2}(\Delta_{\Theta^{-1}} \tau_\mathrm{d}\xi)^2}\cos(\varphi)\cos(\Omega_+\Delta_{\Theta^{-1}} \tau_\mathrm{d}/2)
\\
S_{\mathrm{HOM}}=&e^{-\Omega_-^2/(4\xi^2)}e^{-(\Delta_{\Theta^{-1}} \tau_\mathrm{d}\xi)^2}
\end{align}
\end{subequations}
 with $\mathrm{X=MZ,HOM}$ where $\Omega_\pm=\Omega_2\pm\Omega_1$.  In the limit $\Omega_-\gg \xi$ we have $\lim_{\Omega_-/\xi\rightarrow\infty}S_{\mathrm{X}}=0$, and the interference patterns (\ref{EQ:IP}) simplify to
\begin{subequations}
\label{EQ:AIP_}
\begin{align}
P_c^{\mathrm{MZ}}=&  e^{-\frac{1}{2}(\Delta_{\Theta^{-1}} \tau_\mathrm{d}\xi)^2}\cos(\Omega_+\Delta_{\Theta^{-1}} \tau_\mathrm{d}/2)\cos(\Omega_-\Delta_{\Theta^{-1}} \tau_\mathrm{d}/2),\label{EQ:AIPMZ}
\\
P_c^{\mathrm{HOM}}=&e^{-(\Delta_{\Theta^{-1}} \tau_\mathrm{d}\xi)^2}\cos(\Omega_-\Delta_{\Theta^{-1}} \tau_\mathrm{d}-\varphi),\label{EQ:AIPHOM}
\end{align}
\end{subequations}
which in the limit of vanishing frequency separation, i.e. $\Omega_1=\Omega_2=\Omega$, reproduce the well known textbook result of the MZ-interference pattern $P_c^{\mathrm{MZ}}=\cos(\Omega\Delta \tau)$ (for a monochromatic wave, i.e. $\xi=0$) and the seminal result $P_c^{\mathrm{HOM}}=e^{-(\Delta_{\Theta^{-1}}\tau\xi)^2}$ known as the \textit{HOM-Dip} \citep{HOM}, when $\Delta \tau$ is identified with $\Delta_{\Theta^{-1}} \tau_\mathrm{d}$. In the narrow bandwidth limit $\Delta_{\Theta^{-1}} \tau_\mathrm{d}\xi\ll1$, the interference patterns simplify to
\begin{subequations}
\label{EQ:AIPR_}
\begin{align}
P_c^{\mathrm{MZ}}=& \cos(\Omega_+\Delta_{\Theta^{-1}} \tau_\mathrm{d}/2)\cos(\Omega_-\Delta_{\Theta^{-1}} \tau_\mathrm{d}/2),\label{EQ:AIPMZ_2}
\\
P_c^{\mathrm{HOM}}=&\cos(\Omega_-\Delta_{\Theta^{-1}} \tau_\mathrm{d} -\varphi)\label{EQ:AIPHOM_2}\,.
\end{align}
\end{subequations}

}

{
\section{Mixed states, partial traces and identification of entanglement in second quantisation}\label{sec:MixDecoh}
After having derived the measurement statistics of the considered MZ- and HOM-experiments employing QMems and optical delay lines in the two previous appendix sections, this chapter is dedicated to take a closer look at the physical interpretation of these measurement statistics, where we put particular emphasis on the decoherence of the spatial correlation that arises as a consequence of EDs due to gravity. Note that the measurement statistics in any case solely depend on the diagonal elements of the reduced density matrices, as seen in Eq. \eqref{eq:diagentriesredrho} from the main text. However, in order to characterize the degree of coherence of a quantum (sub) state, it is essential to study the off-diagonal elements of the considered density matrices, the elements carrying information about the quantum phases.

Therefore, we first characterize mixed states quantitatively and derive the partial trace in second quantization. Concrete examples are provided by specializing our results to the here discussed MZ- and HOM-experiments, for which we show how the coherence of the spatial substate of the considered photons is lost.

In analogy to our definition of pure single- and two-photon states \eqref{EQ:SPS} and \eqref{EQ:TPS} that are
{
\begin{subequations}
\label{EQ:APS}
\begin{align}
\ket{\psi_{\mathrm{1}}}=&\sum_\sigma\int d\omega\, \psi_{\sigma}(\omega)\hat{a}^\dagger_{\sigma \omega}\ket{0},    \label{EQ:ASPS}
\\
\ket{\psi_{\mathrm{2}}}=& \frac{1}{\sqrt{2}} \sum_{\sigma_1,\sigma_2}\int d\omega_1 d\omega_2\, \psi_{\sigma_1\sigma_2}(\omega_1,\omega_2)\hat{a}^\dagger_{\sigma_1 \omega_1}\hat{a}^\dagger_{\sigma_2 \omega_2}\ket{0},    \label{EQ:ATPS}
\end{align}
\end{subequations}
}
we define mixed single- and two-photon states as
{
\begin{subequations}
\label{EQ:AMS}
\begin{align}
{\hat{\rho}}_{\mathrm{1}}=&\sum_{\sigma,\bar{\sigma}}\int d\omega\, d\bar{\omega}\, \rho_{\sigma\bar{\sigma}}(\omega,\bar{\omega})\,\hat{a}^\dagger_{\sigma \omega}\ket{0}\bra{0}\hat{a}_{\bar{\sigma} \bar{\omega}},    \label{EQ:ASMS}
\\
{\hat{\rho}}_{\mathrm{2}}=& \frac{1}{2}\sum_{\substack{\sigma_1, \sigma_2, \\ \bar{\sigma}_1,\bar{\sigma}_2}}\int d\omega_1\,d\omega_2\, d\bar{\omega}_1\,d\bar{\omega}_2\, \rho_{\sigma_1\sigma_2\bar{\sigma}_1\bar{\sigma}_2}(\omega_1,\omega_2,\bar{\omega}_1,\bar{\omega}_2)\,\hat{a}^\dagger_{\sigma_1 \omega_1}\hat{a}^\dagger_{\sigma_2 \omega_2}\ket{0}\bra{0}\hat{a}_{\bar{\sigma}_1 \bar{\omega}_1}\hat{a}_{\bar{\sigma}_2 \bar{\omega}_2},    \label{EQ:ATMS}
\end{align}
\end{subequations}
}
which are characterized through the corresponding density matrices $\rho_{\sigma\bar{\sigma}}(\omega,\bar{\omega})$ and $\rho_{\sigma_1\sigma_2\bar{\sigma}_1\bar{\sigma}_2}(\omega_1,\omega_2,\bar{\omega}_1,\bar{\omega}_2)$ rendering the following symmetries
\begin{subequations}
\begin{align}
    \rho_{\sigma\bar{\sigma}}(\omega,\bar{\omega})&=\rho^*_{\bar{\sigma}\sigma}(\bar{\omega},\omega)\label{EQ:ASym1}
    \\
    \rho_{\sigma_1\sigma_2\bar{\sigma}_1\bar{\sigma}_2}(\omega_1,\omega_2,\bar{\omega}_1,\bar{\omega}_2)
    =
    \rho_{\sigma_2\sigma_1\bar{\sigma}_1\bar{\sigma}_2}(\omega_2,\omega_1,\bar{\omega}_1,\bar{\omega}_2)
    &=
    \rho_{\sigma_1\sigma_2\bar{\sigma}_2\bar{\sigma}_1}(\omega_1,\omega_2,\bar{\omega}_2,\bar{\omega}_1)
    =
    \rho^*_{\bar{\sigma}_1\bar{\sigma}_2\sigma_1\sigma_2}(\bar{\omega}_1,\bar{\omega}_2,\omega_1,\omega_2),\label{EQ:ASym2}
\end{align}
\end{subequations}
which can be inferred by use of the canonical commutation relations. 
Physically, the first two equal signs of Eq. \eqref{EQ:ASym2} are a consequence of the requirement for a bosonic two-particle density function to be symmetric under particle exchange, and the last equal signs from Eqs. \eqref{EQ:ASym1} and \eqref{EQ:ASym2} reflect the hermicity of the density operators.
For pure states, the density matrices take the form of dyadic products
\begin{subequations}
\begin{align}
    \rho_{\sigma\bar{\sigma}}(\omega,\bar{\omega})&=\psi_{\sigma}(\omega)\psi^*_{\bar{\sigma}}(\bar{\omega})
    \\
    \rho_{\sigma_1\sigma_2\bar{\sigma}_1\bar{\sigma}_2}(\omega_1,\omega_2,\bar{\omega}_1,\bar{\omega}_2)&=\psi_{\sigma_1\sigma_2}(\omega_1,\omega_2)\psi^*_{\bar{\sigma}_1\bar{\sigma}_2}(\bar{\omega}_1,\bar{\omega}_2).
\end{align}
\end{subequations}

To consider substates of a quantum system, basically any DOFs can be traced out by the \textit{partial trace}. In this article, we trace out certain DOFs of a quantum system shared by all particles, thereby leaving the particle number of the system unchanged. In MZ- and HOM-interference, one is usually interested in the spatial substate of the photons (i.e. at which detector a photon is detected) for the case of frequency insensitive photo detectors.

{ To consistently define the partial trace, we have to define a tensor product structure first. To this end, we assume that the single particle Hilbert space $\mathcal{H}$ can be written consistently as the tensor product of a spatial Hilbert space $\mathcal{H}_\mathrm{s}$ and a frequency Hilbert space $\mathcal{H}_\mathrm{f}$, i.e. $\mathcal{H}=\mathcal{H}_\mathrm{s}\otimes\mathcal{H}_\mathrm{f}$, which implies that there must exist an abstract basis of $\mathcal{H}$ of single particle wave functions that can be written as $\{|\sigma\rangle \otimes|\omega\rangle\}_{\sigma,\omega}$, where $\{|\sigma\rangle\}_{\sigma}$ is a basis of $\mathcal{H}_\mathrm{s}$ and $\{|\omega\rangle\}_{\omega}$ is a basis of $\mathcal{H}_\mathrm{f}$, i.e. spatial degree of freedom and frequency completely decouple. 

From the single particle Hilbert space, we construct the 2-particle Hilbert space as the tensor product, $\mathcal{H}\otimes\mathcal{H}$ which we can identify with $\mathcal{H}_\mathrm{s}\otimes  \mathcal{H}_\mathrm{f}\otimes \mathcal{H}_\mathrm{s} \otimes \mathcal{H}_\mathrm{f}$ via the basis $\{|\sigma\rangle\otimes |\omega\rangle\otimes|\bar\sigma\rangle \otimes|\bar\omega\rangle\}_{\vec\sigma,\vec\omega}$, where $\vec\sigma=(\sigma,\bar\sigma)$ and $\vec\omega=(\omega,\bar{\omega})$. A simple rearrangement enables the identification with $\mathcal{H}_\mathrm{s}^{\otimes2}\otimes\mathcal{H}_\mathrm{f}^{\otimes2}$ with basis states $\{|\sigma,\bar\sigma\rangle\otimes |\omega,\bar\omega\rangle\}_{\vec\sigma,\vec\omega}$, where $|\sigma,\bar\sigma\rangle=|\sigma\rangle\otimes|\bar\sigma\rangle$ and $|\omega,\bar\omega\rangle=|\omega\rangle\otimes|\bar\omega\rangle$. Note that this structure implicitly requires a labelling of the photons with the position in the state vector $|\sigma,\bar\sigma\rangle$ and $|\omega,\bar\omega\rangle$. Since photons are indistinguishable the symmetrization has to be subsequent. Explicitly, we obtain
\begin{equation}
    \hat{a}^\dagger_{\sigma\omega}\hat{a}^\dagger_{\bar\sigma\bar\omega}|0\rangle = \frac{1}{\sqrt{2}}\left(|\sigma,\bar\sigma\rangle\otimes |\omega,\bar\omega\rangle + |\bar\sigma,\sigma\rangle\otimes |\bar\omega,\omega\rangle \right)\,.
\end{equation} 
Therefore, we can write the partial trace of the two-particle density matrix over the frequency sub-space as
\begin{align}
\mathrm{Tr}_f[{\hat{\rho}}_{\mathrm{2}}]=& \frac{1}{2}\sum_{\substack{\sigma_1, \sigma_2, \\ \bar{\sigma}_1,\bar{\sigma}_2}}\int d\omega_1' d\omega_2' d\omega_1 d\omega_2 d\bar{\omega}_1 d\bar{\omega}_2\, \rho_{\sigma_1\sigma_2\bar{\sigma}_1\bar{\sigma}_2}(\omega_1,\omega_2,\bar{\omega}_1,\bar{\omega}_2)\,\langle\omega_1',\omega_2'|\hat{a}^\dagger_{\sigma_1 \omega_1}\hat{a}^\dagger_{\sigma_2 \omega_2}\ket{0}\bra{0}\hat{a}_{\bar{\sigma}_1 \bar{\omega}_1}\hat{a}_{\bar{\sigma}_2 \bar{\omega}_2}|\omega_1',\omega_2'\rangle \nonumber \\
=&   \frac{1}{4}\sum_{\substack{\sigma_1, \sigma_2, \\ \bar{\sigma}_1,\bar{\sigma}_2}}\int d\omega_1' d\omega_2' d\omega_1 d\omega_2 d\bar{\omega}_1 d\bar{\omega}_2\, \rho_{\sigma_1\sigma_2\bar{\sigma}_1\bar{\sigma}_2}(\omega_1,\omega_2,\bar{\omega}_1,\bar{\omega}_2) \times\nonumber \\
&\times \left( |\sigma_1,\sigma_2\rangle\delta(\omega_1'-\omega_1)\delta(\omega_2'-\omega_2) + |\sigma_2,\sigma_1\rangle\delta(\omega_1'-\omega_2)\delta(\omega_2'-\omega_1) \right)\times   \nonumber \\
&\times \left( \langle\bar\sigma_1,\bar\sigma_2|\delta(\omega_1'-\bar\omega_1)\delta(\omega_2'-\bar\omega_2) + \langle\bar\sigma_2,\bar\sigma_1|\delta(\omega_1'-\bar\omega_2)\delta(\omega_2'-\bar\omega_1) \right)  \\
=&  \sum_{\substack{\sigma_1, \sigma_2, \\ \bar{\sigma}_1,\bar{\sigma}_2}}\int d\omega_1' d\omega_2'\, \rho_{\sigma_1\sigma_2\bar{\sigma}_1\bar{\sigma}_2}(\omega_1',\omega_2',\omega_1',\omega_2') |\sigma_1,\sigma_2\rangle\langle\bar\sigma_1,\bar\sigma_2|   \nonumber \\
=: & \sum_{\substack{\sigma_1, \sigma_2, \\ \bar{\sigma}_1,\bar{\sigma}_2}}  \rho_{\sigma_1\sigma_2\bar{\sigma}_1\bar{\sigma}_2} |\sigma_1,\sigma_2\rangle\langle\bar\sigma_1,\bar\sigma_2|  \nonumber \,,   
\end{align}
where we used the symmetries in equation \eqref{EQ:ASym2}. Note that the spatial states $|\sigma_1,\sigma_2\rangle\langle$ inherit the implicit labeling of particles with the position in the state vector. Therefore in general, $|\sigma_1,\sigma_2\rangle$ is not the same state as $|\sigma_2,\sigma_1\rangle$ for $\sigma_1\neq\sigma_2$. This is only the case if the two involved photons are completely indistinguishable, that is, are in the same frequency state. Then, $\hat\rho_2$ is separable in spatial and frequency degrees of freedom and the reduced density matrix is that of a pure state. Accordingly, the reduced density matrix $\rho_{\sigma_1\sigma_2\bar{\sigma}_1\bar{\sigma}_2}$ is not generically symmetric
under the exchange of indices within the pairs $(\sigma_1,\sigma_2)$ and $(\bar\sigma_1,\bar\sigma_2)$.

For the n-particle sector, the reduced density matrix is simply defined as 
\begin{align}
    \rho_{\sigma_1...\sigma_n\bar{\sigma}_1...\bar{\sigma}_n}=\int d\omega_1...\,d\omega_n\, \rho_{\sigma_1...\sigma_n\bar{\sigma}_1...\bar{\sigma}_n}(\omega_1,...,\omega_n,{\omega}_1,...,{\omega}_n).\label{EQ:APTrRw}\,.
\end{align}
A measure of the entanglement between the spatial and frequency degrees of freedom can then be defined through the purity of the reduced density matrix}
\begin{align}
    \mathcal{P}(\rho^\mathrm{E})=\sum_{\substack{\sigma_1,..., \sigma_n, \\ \bar{\sigma}_1,...,\bar{\sigma}_n}}|\rho^\mathrm{E}_{\sigma_1...\sigma_n\bar{\sigma}_1...\bar{\sigma}_n}|^2,
\end{align}
where values below one indicate the presence of entanglement between the internal and external DOFs in case that the original state is pure.

Next we compute the reduced spatial density matrices at the detectors of the MZ- and HOM-experiments which employ QMems, and similar considerations hold for the case of delay lines.

For the MZ- experiment that employs QMems the density matrix at the detectors is equal to
\begin{align}
    \rho_{\sigma\bar{\sigma}}(\omega,\bar{\omega})=\psi^\mathrm{f}_{\sigma}(\omega)\psi^\mathrm{f}_{\bar{\sigma}}(\bar{\omega}),\ \sigma,\bar{\sigma}=\oplus,\ominus\label{EQ:AMZx}
\end{align}
where $\psi^\mathrm{f}_{\sigma}(\omega)$ has to be taken from \eqref{EQ:APF1}, and can be given in vectorial notation as.
Then the computation of the reduced spatial density matrix results in
\begin{align}
    \rho_{\sigma\bar{\sigma}}=\int d\omega \rho_{\sigma\bar{\sigma}}(\omega,\bar{\omega})=
    \frac{1}{2}
    \begin{pmatrix}
        1-\cos(\Omega_+\Delta_{\Theta^{-1}}\tau_\mathrm{s}/2)\cos(\Omega_-\Delta_{\Theta^{-1}} \tau_\mathrm{s}/2)
        & 
        -i\sin(\Omega_+\Delta_{\Theta^{-1}} \tau_\mathrm{s}/2)\cos(\Omega_-\Delta_{\Theta^{-1}} \tau_\mathrm{s}/2)
        \\
        i\sin(\Omega_+\Delta_{\Theta^{-1}} \tau_\mathrm{s}/2)\cos(\Omega_-\Delta_{\Theta^{-1}} \tau_\mathrm{s}/2)
         &
    1+\cos(\Omega_+\Delta_{\Theta^{-1}} \tau_\mathrm{s}/2)\cos(\Omega_-\Delta_{\Theta^{-1}} \tau_\mathrm{s}/2)
    \end{pmatrix},\label{EQ:ARRs1}
\end{align}
where we have directly taken the large bandwidth limit $\Omega_-\gg \xi$, and have set
\begin{align}
    \int d\omega \,g_{\Omega_i\xi}(\omega)g_{\Omega_j\xi}(\omega)=\delta_{ij},\ N_\mathrm{MZ}=1/\sqrt{2}
\end{align}
for better readability of Eq. \eqref{EQ:ARRs1}. It follows immediately that the purity of the reduced density matrix of the spatial substate \eqref{EQ:ARRs1} in the considered MZ-experiment equals
\begin{align}
    \mathcal{P}_\mathrm{MZ}(\tau_\mathrm{s})=\sum_{\sigma,\bar{\sigma}=\oplus,\ominus}|\rho_{\sigma\bar{\sigma}}|^2=\frac{1}{4}(3+\cos(\Omega_-\Delta_{\Theta^{-1}} \tau_\mathrm{s}))
,
\end{align}
which reaches its minimum value of $\mathcal{P}(\tau^{\mathrm{MZ}}_\mathrm{s,ent})=1/2$ at
\begin{align}
    \tau^{\mathrm{MZ}}_\mathrm{s,ent}=\frac{\pi}{\Omega_-\Delta_{\Theta^{-1}}},
\end{align}
which we call \textit{entangling time} since at this time the spatial substate \eqref{EQ:ARRs1} is apparently mixed, although the entire photonic state (\ref{EQ:AMZx}) is obviously pure, i.e. that the external DOF of the considered photon must be entangled with its internal DOF, the frequency. 

For the HOM- experiments that employ QMems the density matrix at the detectors is equal to
\begin{align}
\rho_{\sigma_1\sigma_2\bar{\sigma}_1\bar{\sigma}_2}(\omega_1,\omega_2,\bar{\omega}_1,\bar{\omega}_2)=\psi^\mathrm{f}_{\sigma_1\sigma_2}(\omega_1,\omega_2)\psi^\mathrm{f}_{\bar{\sigma}_1\bar{\sigma}_2}(\bar{\omega}_1,\bar{\omega}_2),
\end{align}
where $\psi^\mathrm{f}_{\sigma_1\sigma_2}(\omega_1,\omega_2)$ has to be taken from \eqref{EQ:APF2}. Then the computation of the
reduced spatial density matrix results in
\begin{align} \rho_{\sigma_1\sigma_2\bar{\sigma}_1\bar{\sigma}_2}
=
\frac{1}{4}
\begin{pmatrix}
    1+\cos(p) & 0 & 0 & -1-\cos(p)\\
    0 & 1-\cos(p) & -1+\cos(p) & 0\\
    0 & -1+\cos(p) & 1-\cos(p) & 0\\
    -1-\cos(p) & 0 & 0 & 1+\cos(p)
\end{pmatrix},\label{EQ:Add}
\end{align}
where $p=\Omega_-\Delta_{\Theta^{-1}} \tau_\mathrm{s}-\varphi$. In Eq. (\ref{EQ:Add}) the row and column numbering corresponds to $\{\oplus\oplus,\oplus\ominus,\ominus\oplus,\ominus\ominus\}$, and we again employed the large bandwidth limit for better readability. The purity of the spatial substate in this case results in
\begin{align}
        \mathcal{P}_\mathrm{HOM}(\tau_\mathrm{s})=\sum_{\sigma_1,\bar{\sigma}_1,\sigma_2,\bar{\sigma}_2=\oplus,\ominus}|\rho_{\sigma_1\sigma_2\bar{\sigma}_1\bar{\sigma}_2}|^2=\frac{1}{4}(3+\cos(2\Omega_-\Delta_{\Theta^{-1}} \tau_\mathrm{s}-2\varphi))
\end{align}
which for $\varphi=0$ reaches its minimum value of $\mathcal{P}(\tau^{\mathrm{HOM}}_\mathrm{s,ent})=1/2$ at
\begin{align}
    \tau^{(2)}_\mathrm{s,ent}=\frac{\pi}{2\Omega_-\Delta_{\Theta^{-1}}}.
\end{align}

We want to add that, from an information theoretic perspective, it is more conventional to quantify the mixedness of a quantum state by the linear entropy $\mathcal{S}=1-\mathcal{P}$ instead of the purity $\mathcal{P}$, where now  $\mathcal{S}=0$ corresponds to pure states, i.e. values $\mathcal{S}>0$ of reduced substates of a pure quantum state indicate the presence of entanglement. For the here discussed MZ- and HOM-experiments, the linear entropy results in
\begin{subequations}
\begin{align}
    \mathcal{S}_\mathrm{MZ}(\tau_\mathrm{s})=&1-\mathcal{P}_\mathrm{MZ}(\tau_\mathrm{s})=\frac{1}{2}\sin^2\left(\Omega_-\Delta_{\Theta^{-1}} \tau_\mathrm{s}/2\right)\\
    \mathcal{S}_\mathrm{HOM}(\tau_\mathrm{s})=&1-\mathcal{P}_\mathrm{HOM}(\tau_\mathrm{s})=\frac{1}{2}\sin^2\left(\Omega_-\Delta_{\Theta^{-1}} \tau_\mathrm{s}-\varphi\right),
\end{align}
\end{subequations}
which stands in perfect analogy to the corresponding results (\ref{EQ:1Q}) and (\ref{EQ:A2Q}), which describe the gravitationally induced dynamics of entanglement between the internal and external DOFs of qubit clock states. 

Finally, we want to investigate the entanglement contained in the spatial sub-state (\ref{EQ:Add}), i.e. the mutual spatial entanglement shared among the two photons, which can be explored by frequency insensitive detectors in the considered HOM-experiment. The density matrix (\ref{EQ:Add}) characterizes a two-qubit state, where one qubit is encoded in the spatial DOF of one photon and the other qubit is encoded in the spatial DOF of the other photon. In such a system the presence of entanglement can be unambiguously certified by positive values of the negativity, which is equal to the negative sum of the negative eigenvalues of the partially transposed density matrix of (\ref{EQ:Add}), which is 
\begin{align} \rho^{\mathrm{PT}}_{\sigma_1\sigma_2\bar{\sigma}_1\bar{\sigma}_2}
=
\frac{1}{4}
\begin{pmatrix}
    1+\cos(p) & 0 & 0 & -1+\cos(p)\\
    0 & 1-\cos(p) & -1-\cos(p) & 0\\
    0 & -1-\cos(p) & 1-\cos(p) & 0\\
    -1+\cos(p) & 0 & 0 & 1+\cos(p)
\end{pmatrix},\label{EQ:AddPT}
\end{align}
and featuring the eigenvalues $\lambda_{\mathrm{PT}}=1/2(1,1,\cos(p),-\cos(p))$. Thus, the associated negativity results in
\begin{align}
\mathcal{N}_\mathrm{HOM}=-\sum_{\lambda_{\mathrm{PT}}<0} \lambda_{\mathrm{PT}}=|\cos\left(\Omega_-\Delta_{\Theta^{-1}} \tau_\mathrm{s}-\varphi\right)|/2, 
\end{align}
thereby establishing the analogy to the result (\ref{EQ:Neg}), which characterizes the mutual spatial entanglement of two particles subject to EDs due to gravity from Appendix \ref{sec:A1}. 

In conclusion, we can simplify the expressions for the linear entropy and negativity by expressing them in terms of the quantities $P_c^{\mathrm{HOM}}=\cos(\Omega_-\Delta_{\Theta^{-1}} \tau_{\mathrm{s}}-\varphi)$ and $\mathcal{V}=\cos(\Omega_-\Delta_{\Theta^{-1}} \tau_{\mathrm{s}}/2)$. We obtain
\begin{subequations}
\begin{align}
    \mathcal{S}_\mathrm{MZ}=&\frac{1}{2}\left(1-\mathcal{V}^2\right)
    \\
    \mathcal{S}_\mathrm{HOM}=&\frac{1}{2}\left(1-\left(P_c^{\mathrm{HOM}}\right)^2\right),
    \\
    \mathcal{N}_\mathrm{HOM}=&\frac{|P_c^{\mathrm{HOM}}|}{2}.
\end{align}
\end{subequations}

}

\section{Effect of longitudinal Doppler 
shift}
\label{app:longredshift}

In this appendix, we estimate the effect of the longitudinal Doppler shift between two satellites that are moving on circular orbits. The expression for the Doppler shift to leading order (Newtonian) is $\Delta v/c$, where $\Delta v_\mathrm{long}$ is the relative velocity of the two satellites. In the following, we assume that both orbits lie in the equatorial plane, and at $t=0$, the positions of the two satellites are aligned with the center of Earth which implies that $\Delta v(t=0)=0$. We find for the distance between the two satellites at $t>0$
\begin{equation}
    \Delta R(t) = \sqrt{r_1^2+r_2^2 - 2r_1r_2 \cos(\omega_1-\omega_2)t}
\end{equation}
and for the relative velocity 
\begin{equation}
    \Delta v_\mathrm{long}(t) = \frac{r_1r_2(\omega_1-\omega_2)\sin(\omega_1-\omega_2)t}{\sqrt{r_1^2+r_2^2 - 2r_1r_2 \cos(\omega_1-\omega_2)t}}\,,
\end{equation}
where $\omega_{1/2}=\sqrt{GM/r_{1/2}^3}$ and $r_{1/2}$ are the orbital angular frequencies and radii of the two satellites, respectively. 

Let us assume that we have two circular orbits, one at 10000 km above the Earth's surface (16371 km from the center) and one at a geosynchronus orbit. Let us further assume that the pulse from the upper arm arrives at $U$ at $t=0$. Then, at $t=\tau_d=8.5\,\mathrm{km}/c \sim 3\times 10^{-5}\,\mathrm{s}$ when the pulse from the lower arm arrives at the beam splitter, there will be a red-shift  of $z_\mathrm{Doppler}(\tau_d)=\Delta v_\mathrm{long}(\tau_d)/c \sim 1\times 10^{-13}$ and a phase difference due to the difference in travel distance of the photons of $\Delta R(\tau_d)/\lambda \sim 4\times 10^{-4}$. So it seems that the longitudinal Doppler effect can indeed be neglected if the timing of the experiment is precise enough.


\section{Estimation of the required temporal resolution}
\label{sec:estimation}

In the following we estimate the resolution of the optical delays and the storage time of the QMems, which is required in order to infer that the relativistic redshift is the origin of the loss of spatial correlation between the photons in the HOM-experiment considered in Fig. \ref{fig:times} from the main text, and not just appropriate (noisy) delay settings as discussed in the main text. In Eq. \eqref{EQ:IPHOM} the argument of the cosine function is 
\begin{align}
   C(\tau_\mathrm{s}):=\Delta_{\Theta^{-1}}\Omega_-\tau_\mathrm{s}, \label{EQ:AT1}
\end{align}
where here and in the following the subscript "s" for storage time can be replaced the subscript "d" for delay in order to obtain the corresponding analysis for optical delays.
Note that the result (\ref{EQ:AT1}) is based on the assumption that both storage times of the QMems in the two distinct arms of the interferometer  $\tau_{U,\mathrm{s}}$ and $\tau_{L,\mathrm{s}}$ are locally adjusted to precisely the same value $\tau_{U,\mathrm{s}}=\tau_{L,\mathrm{s}}=\tau_{\mathrm{s}}$ with infinite precision. However, in practice the storage time in each arm can only be adjusted to a finite precision, the absolute resolution error $\tau^\mathrm{res}_\mathrm{s}$. 
In the general case, that is, $\tau_{U,\mathrm{s}} \neq \tau_{L,\mathrm{s}}$ Eq. \eqref{EQ:AT1} generalizes to
\begin{align}
    C(\tau_{U,\mathrm{s}},\tau_{L,\mathrm{s}})=\Omega_-\left(\tau_{U,\mathrm{s}}/\Theta_U-\tau_{L,\mathrm{s}}/\Theta_L\right),\label{EQ:AC}
\end{align}
We are interested in the error of $C$ (which we call $\Delta C$), which is caused by the uncertainties of the storage times of the QMems  $\tau^\mathrm{res}_\mathrm{s}$ and find from the standard error estimation
\begin{align}
    \Delta C=&\left(\left|\frac{\partial C}{\partial \tau_{U,\mathrm{s}}}\right|+\left|\frac{\partial C}{\partial \tau_{L,\mathrm{s}}}\right|\right)\tau^\mathrm{res}_{\mathrm{s}}
    =
    \Omega_-\left(1/\Theta_U+1/\Theta_L\right)\tau^\mathrm{res}_{\mathrm{s}}
    =
     \Omega_-\left(\frac{1}{1+z_{sU}}+\frac{1}{1+z_{sL}}\right)\tau^\mathrm{res}_{\mathrm{s}}
    \approx 2\Omega_-\tau^\mathrm{res}_{\mathrm{s}}.
\end{align}
In order to exclude the possibility that the finite resolution effects are the reason for the loss of spatial correlation in the considered HOM-experiment (and not relativistic effects) one has to require  $\Delta C \ll \pi/2$ (since we are interested in the parameters, which render $C=\pi/2$). A necessary condition for this are the restrictions on the resolution of the optical delays and QMems
\begin{subequations}
\label{EQ:ARES}
\begin{align}
    &\tau^\mathrm{res}_\mathrm{s} \ll \tau^\mathrm{reslim}_\mathrm{s}=\frac{\pi}{4\Omega_-}. 
\end{align}
\end{subequations}
which we displayed as the top axis in Fig. \ref{fig:times} of the main text. The superscript "reslim" stands for resolution limit.

Vice versa Eq. \eqref{EQ:ARES} yields the maximum frequency separations for a given temporal resolution and $\tau^\mathrm{res}_\mathrm{s}$, that is
\begin{subequations}
\begin{align}
    \Omega_- \ll \Omega_-^\mathrm{lim} =&\frac{\pi}{4\tau^\mathrm{res}_\mathrm{s}}.
\end{align}
\end{subequations}

Finally, we want to remark that the above requirement on the temporal resolution is identical to the following condition on the relative temporal accuracy at the entanglement time 
\begin{align}
    \frac{\tau^\mathrm{res}_\mathrm{s}}{\tau^{\mathrm{HOM}}_\mathrm{s,ent}}\ll \frac{\tau^\mathrm{reslim}_\mathrm{s}}{\tau^{\mathrm{HOM}}_\mathrm{s,ent}}=\frac{\Delta_{\Theta^{-1}}}{2},
\end{align}
for the case of HOM interference, and an analogous expression is obtained for MZ interference.

\section{Different QMem combinations}\label{sec:ATab}

Below, the differences between the acceptance frequencies of the various QMem combinations which are discussed in chapter \ref{sec:expimp} in the main text are seen. As discussed in the main text, inhomogenous broadening of the REID materials would enable frequency multiplexing with $\Omega_-\sim1-10$~GHz. And in fact, storing frequency superposition states with $\Omega_-$ as small as $\sim$10~MHz is possible with these system but then the required storage time will be out of reach with such QMems.
\begin{center}
\begin{tabular}{ | m{0.5cm} | m{3.5cm}| m{3.5cm} |  m{3.5cm} | m{3.5cm}  |} 
  \hline
   & Rb  (377.1 THz (795\,nm) ~\cite{Dudin2013}) & Cs (335.3 THz (894\,nm)~\cite{Katz2018}) & Pr (494.7 THz (606\,nm) ~\cite{Gundogan2015}) & Eu (517.9 THz (579\,nm) ~\cite{Ortu2022}) \\
 \hline
  Rb &  & 41.8 & 117.6 & 140.9\\ 
 \hline
 Cs&  &  & 159.6 & 182.6\\ 
  \hline
  Pr &  &  & $10^{-3} - 10^{-2}$ & 23.2 \\ 
  \hline
 Eu &  &  &  & $10^{-3} - 10^{-2}$ \\ 
  \hline
\end{tabular}
\vspace{10pt}\\
Table 1: The achievable $\Omega_-$ for different QMem combinations. If not stated differently, all quantities in the table are given in units of THz.

\end{center}

\end{document}